\documentclass[conference]{IEEEtran}
\IEEEoverridecommandlockouts
\usepackage{graphicx}      % For including graphics
\graphicspath{{fig/}} 
\usepackage{listings}      % For code listings
\usepackage{xcolor}        % For colors
\usepackage{booktabs}
\usepackage{xcolor}        % For colors
\usepackage{hyperref}
\usepackage{algorithm}
\usepackage{algorithmic}
\usepackage{balance}

\usepackage{listings}
\usepackage{caption} % For \captionof
\usepackage{caption}    % For captions outside of floats
\usepackage{amsmath}    % For mathematical symbols
\usepackage{xcolor}     % For colored syntax highlighting
\usepackage{amssymb}
\usepackage{amsmath}

\usepackage{listings}   % For code listings
\usepackage{algorithm}  % For the algorithm environment
\usepackage{caption}    % For captions outside of floats
\usepackage{xcolor}     % For colored syntax highlighting

\lstdefinelanguage{JavaScript}{
	keywords={typeof, new, true, false, catch, function, return, null, catch, switch, var, if, in, while, do, else, case, break},
	keywordstyle=\color{blue}\bfseries,
	ndkeywords={class, export, boolean, throw, implements, import, this},
	ndkeywordstyle=\color{orange}\bfseries,
	identifierstyle=\color{black},
	sensitive=false,
	comment=[l]{//},
	morecomment=[s]{/*}{*/},
	commentstyle=\color{gray}\ttfamily,
	stringstyle=\color{red}\ttfamily,
	morestring=[b]',
	morestring=[b]"
}

\author{\IEEEauthorblockN{Mordechai Guri, Dor Fibert}
	\IEEEauthorblockA{\textit{Ben-Gurion University of the Negev} \\
		\textit{Beer Sheva, Israel} \\
		gurim@post.bgu.ac.il ; fibert@post.bgu.ac.il \\ \url{https://www.covertchannels.com/}}
}

\begin{document}


\title{Browser Fingerprinting Using WebAssembly} 
\maketitle

\begin{abstract}
Web client fingerprinting has become a widely used technique for uniquely identifying users, browsers, operating systems, and devices with high accuracy. While it is beneficial for applications such as fraud detection and personalized experiences, it also raises privacy concerns by enabling persistent tracking and detailed user profiling. This paper introduces an advanced fingerprinting method using WebAssembly (Wasm)—a low-level programming language that offers near-native execution speed in modern web browsers. With broad support across major browsers and growing adoption, WebAssembly provides a strong foundation for developing more effective fingerprinting methods.

In this work, we present a new approach that leverages WebAssembly’s computational capabilities to identify returning devices—such as smartphones, tablets, laptops, and desktops—across different browsing sessions. Our method uses subtle differences in the WebAssembly JavaScript API implementation to distinguish between Chromium-based browsers like Google Chrome and Microsoft Edge, even when identifiers such as the User-Agent are completely spoofed, achieving a false-positive rate of less than 1\%. The fingerprint is generated using a combination of CPU-bound operations, memory tasks, and I/O activities to capture unique browser behaviors. We validate this approach on a variety of platforms, including Intel, AMD, and ARM CPUs, operating systems such as Windows, macOS, Android, and iOS, and in environments like VMWare, KVM, and VirtualBox. Extensive evaluation shows that WebAssembly-based fingerprinting significantly improves identification accuracy. We also propose mitigation strategies to reduce the privacy risks associated with this method, which could be integrated into future browser designs to better protect user privacy.

%By addressing the growing concerns around user tracking, this work lays the groundwork for more privacy-aware web technologies while maintaining the effectiveness of web client identification.
\end{abstract}



%\keywords{Browser Fingerprinting, Device Fingerprinting, WebAssembly, Privacy, Wasm, Web Clients}

%\maketitle

\section{Introduction}
Over the years, a diverse set of fingerprinting techniques has been developed to uniquely identify web pages \cite{Sirinam2018, Wang2014, Korczynski2014}, web browsers \cite{Cao2017, Nikiforakis2013, Fifield2015a}, users \cite{Das2017, Nikiforakis2013, Sanchez-Rola2018}, operating systems \cite{nmap, Shamsi2014, Nikiforakis2013}, and even specific devices \cite{Miettinen2017, Kumar2018, Zuo2019}. While these techniques serve legitimate purposes, such as enhancing security through advanced authentication mechanisms, they also pose significant privacy risks. For instance, device fingerprinting can improve security by prompting additional authentication steps when an application is accessed from a new device, thereby creating an additional barrier against impersonation attacks. At the same time, adversaries can misuse these methods to profile a target's browser or operating system, potentially exposing vulnerabilities and enabling the deployment of targeted exploits.

Traditional fingerprinting techniques often rely on measuring device performance through task execution and analyzing timing results \cite{Sanchez-Rola2018}. When implemented in a web environment, these methods face challenges due to the inherent overhead and inconsistent performance of JavaScript—the primary language for client-side code execution. Substantial overhead and unpredictable execution times, caused by just-in-time (JIT) compilation \cite{Selakovic2016}, make it difficult to achieve reliable fingerprinting results.

To address these limitations, this research investigates the potential of using WebAssembly as an alternative platform for web client fingerprinting. WebAssembly, a low-level language designed for efficient execution within browsers, offers near-native performance and consistent execution times, making it an ideal candidate for developing more accurate and reliable fingerprinting methods.

\subsection{WebAssembly}
WebAssembly is a low-level bytecode language designed to execute within a secure sandbox environment in web browsers, providing near-native performance. As of this study, WebAssembly is supported across all major browsers and is maintained through a collaborative effort by Google, Microsoft, Mozilla, Apple, and the W3C \cite{Haas2017}. Primarily serving as a compilation target for low-level languages like C, C++, and Rust, WebAssembly enables developers to write high-performance applications that execute with predictable timing and minimal overhead in web browsers—attributes that are crucial for developing precise and reliable fingerprinting techniques.

\subsection{Contribution}
This work explores the use of WebAssembly to develop more precise browser fingerprinting methods by harnessing its near-native execution speed and exploiting variations in WebAssembly implementations across different CPU architectures, browsers, and operating systems. By capitalizing on these attributes, we address the performance inconsistencies inherent in JavaScript-based fingerprinting and introduce more reliable techniques for web client identification. To mitigate the potential privacy risks posed by these methods, we propose strategies aimed at preserving user privacy, thereby ensuring a balanced approach between accurate browser identification and maintaining user anonymity.

\section{Background and Related Work}
This section reviews the literature related to fingerprinting techniques. First, we describe the two main approaches: passive and active fingerprinting. Next, we explore various types of fingerprinting and their implementations in previous research. Finally, we examine different use-cases of fingerprinting techniques.

\subsection{Passive and Active Fingerprinting}
Passive fingerprinting involves characterizing a target by analyzing its network traffic without direct interaction. Since this technique only requires monitoring network traffic, it can be used to fingerprint a wide range of devices, including personal computers (PCs), mobile phones, and Internet of Things (IoT) devices. An attacker can employ passive fingerprinting through various methods, such as executing a Man-In-The-Middle (MITM) attack \cite{mallik2019man} or eavesdropping on wireless networks \cite{mateti2006hacking}. The primary advantage of passive fingerprinting is its stealth—because the attacker only observes and does not send any packets, it is extremely difficult for intrusion detection systems (IDS) to detect the activity. Previous research has attempted to counter this approach using traffic normalizers \cite{malan2000transport}, which remove identifying information from network packets. However, these defenses are insufficient against newer fingerprinting methods that utilize characteristics such as packet frequency, size, and burst patterns \cite{Sirinam2018, Wang2014}. One limitation of passive fingerprinting is its dependence on the target's network activity, which can result in delays if the target is not actively using the network or is not communicating in a predictable manner, such as issuing a DHCP request \cite{Papapanagiotou2012} or accessing a specific website \cite{Sirinam2018, Wang2014, Korczynski2014}.

In contrast, active fingerprinting involves directly interacting with the target by sending network requests or running code on the device, such as JavaScript in a web browser. Network-based active fingerprinting typically uses protocols like ICMP, TCP, and SNMP to elicit responses that reveal specific characteristics of the target, such as analyzing ICMP headers \cite{nmap, arkin2002remote} or IPv6 headers \cite{Beck2007, Fifield2015b}. Active fingerprinting can also be performed by executing code on the target device, either through a native application on a PC or smartphone \cite{Kumar2018, Sanchez-Rola2018, Quynh2010, Kurtz2015} or via JavaScript scripts in a web browser \cite{Zhang2019, Das2017, Olejnik2016, Cao2017}. However, active fingerprinting is more detectable because IDS, antivirus software, and browser plugins can identify and block suspicious network activity.

\subsection{Overview of Fingerprinting Techniques}
This section provides an overview of common fingerprinting techniques and their applications, highlighting how they have been employed in prior research. Table \ref{table:ScientificBackGroundFingerprintUsages} summarizes the methods used for each fingerprinting type across various research studies.

\begin{table}[h]
	\centering
	\caption{Summary of fingerprinting methods used in various research studies.}
	\label{table:ScientificBackGroundFingerprintUsages}
	\setlength{\tabcolsep}{3pt} % Adjust column separation for IEEE single column format
	\renewcommand{\arraystretch}{1.1} % Adjust line spacing for better readability
	\begin{tabular}{p{0.22\linewidth} p{0.28\linewidth} p{0.40\linewidth}}
		\toprule
		\textbf{Fingerprint Type} & \textbf{Method} & \textbf{References} \\
		\midrule
		\textbf{Operating System (OS)} & Network (Active) & \cite{nmap, arkin2002remote, Beck2007, Fifield2015b} \\
		& Network (Passive) & \cite{Shamsi2014, Aksoy2017, Matsunaka2013, Lastovicka2018, Lastovicka2020, Chen2014, Papapanagiotou2012} \\
		& JavaScript & \cite{Nikiforakis2013, Fifield2015a, Mowery2012, Mowery2011, Schwarz2019, Queiroz2019} \\
		\cmidrule{1-3}
		\textbf{VM Presence} & Network (Active) & \cite{Franklin2008} \\
		& JavaScript & \cite{Ho2014} \\
		\cmidrule{1-3}
		\textbf{VM and OS Detection} & Low-Level Hardware & \cite{Quynh2010} \\
		& Memory & \cite{Gu2012, Owens2011} \\
		\cmidrule{1-3}
		\textbf{Device} & Network (Passive) & \cite{Miettinen2017, Papapanagiotou2012, Husak2016} \\
		& Native Code & \cite{Kumar2018, Sanchez-Rola2018, Zhou2014, Das2014, Kurtz2015, Chen2020} \\
		& Bluetooth & \cite{Zuo2019} \\
		& JavaScript & \cite{Sanchez-Rola2018, Zhang2019, Das2017, Olejnik2016, Cao2017, Queiroz2019} \\
		& Acoustic & \cite{Zhou2014, Das2014} \\
		\cmidrule{1-3}
		\textbf{Browser} & Network (Passive) & \cite{Husak2016} \\
		& JavaScript & \cite{Cao2017, Nikiforakis2013, Fifield2015a, Mowery2012, Mowery2011, Schwarz2019, Queiroz2019} \\
		& CSS & \cite{Takei2015} \\
		\cmidrule{1-3}
		\textbf{Website} & Network (Passive) & \cite{Sirinam2018, Wang2014, Korczynski2014} \\
		& JavaScript & \cite{Shusterman2018} \\
		\bottomrule
	\end{tabular}
\end{table}

\noindent \textbf{Operating System (OS) Fingerprinting:} OS fingerprinting identifies the type and version of a device's operating system. This information is valuable for network inventory management, vulnerability assessment, and patch management. Tools like \textit{nmap} \cite{nmap} and \textit{xprobe} \cite{arkin2002remote} utilize responses from ICMP requests and the TCP/IP stack for active OS fingerprinting. Beck et al. \cite{Beck2007} developed a technique that uses the IPv6 neighbor discovery protocol (NDP) for active OS detection, while Shamsi et al. \cite{Shamsi2014} introduced a single-packet OS fingerprinting method based on SYN-ACK TCP segment characteristics. Chen et al. \cite{Chen2014} further refined OS fingerprinting by analyzing multiple features of TCP/IP headers, and Zuo et al. \cite{Zuo2019} extended fingerprinting capabilities to Bluetooth Low Energy (BLE) devices using static UUIDs.

\noindent \textbf{Virtual Machine (VM) Fingerprinting:} This technique detects the presence of virtualization environments. For example, Franklin et al. \cite{Franklin2008} suggested a method for identifying virtual machine managers over the internet using fuzzy benchmarking, while Ho et al. \cite{Ho2014} leveraged timing variations in standard browser operations to detect emulated environments.

\noindent \textbf{Device Fingerprinting:} Device fingerprinting identifies the type or model of a device. Papapanagiotou et al. \cite{Papapanagiotou2012} fingerprinted wireless devices using DHCP request headers, such as Host-Name and Vendor-Class. Miettinen et al. \cite{Miettinen2017} introduced \textit{IoT SENTINEL}, a system for automatically identifying and securing IoT devices. Das et al. \cite{Das2017} explored using motion sensors like accelerometers and gyroscopes for device fingerprinting, and Olejnik et al. \cite{Olejnik2016} examined the privacy implications of the Battery Status API for identifying returning devices.

\noindent \textbf{Browser Fingerprinting:} Browser fingerprinting identifies a client’s browser type and version. Mowery et al. \cite{Mowery2011} used 29 JavaScript performance tests to fingerprint browsers, operating systems, and microarchitectures. Fifield and Egelman \cite{Fifield2015a} proposed a method using on-screen font glyph sizes, while Cao et al. \cite{Cao2017} utilized GPU (via WebGL) and audio stack properties (\texttt{AudioContext}) for cross-browser fingerprinting. Husák et al. \cite{Husak2016} estimated the User-Agent of a client during HTTPS communication by analyzing the SSL/TLS handshake.

\noindent \textbf{Website Fingerprinting:} Website fingerprinting analyzes network traffic patterns to identify access to specific web pages, even when encrypted. For example, Wang et al. \cite{Wang2014} demonstrated an attack using a KNN classifier to exploit multimodal properties of web pages. Sirinam et al. \cite{Sirinam2018} introduced \textit{Deep-Fingerprinting}, a method targeting the Tor browser using packet size frequencies, transmission times, and burst patterns.

\subsection{Use Cases for Fingerprinting}
Fingerprinting techniques have a wide range of applications, serving both legitimate and malicious purposes. Observers and attackers alike can utilize fingerprinting to identify devices or application instances based on information elements communicated to them.

Fingerprinting techniques are often used for privacy-compromising purposes, such as tracking users and undermining their anonymity \cite{Das2014, Zhou2014, Sirinam2018}. Another key motivation is vulnerability detection, where fingerprinting methods help identify weak spots in a network or device, enabling administrators to apply patches before an attack \cite{Zuo2019}. For network administrators, fingerprinting supports inventory management by providing insights into the devices connected to a network and their configurations \cite{Husak2016, Sanchez-Rola2018}. Fingerprinting can also be useful in identifying returning clients, especially since traditional tracking methods like cookies are often blocked or deleted \cite{Das2017, Sanchez-Rola2018}. Moreover, fingerprinting strengthens authentication by differentiating between legitimate users and imposters based on unique device characteristics \cite{Das2017, Sanchez-Rola2018, Kumar2018}. Conversely, attackers can use fingerprinting to gather specific information about their target devices for deploying tailored exploits. In the realm of cybersecurity, fingerprinting is applied to evade honeypots by distinguishing between real devices and virtual environments \cite{Ho2014}. Additionally, fingerprinting can be used for software license binding to ensure that software is used only on authorized devices \cite{Sanchez-Rola2018, Ho2014}.

\section{WebAssembly Fingerprinting}
\begin{figure}[t]
	\centering
	\includegraphics[width=1\columnwidth]{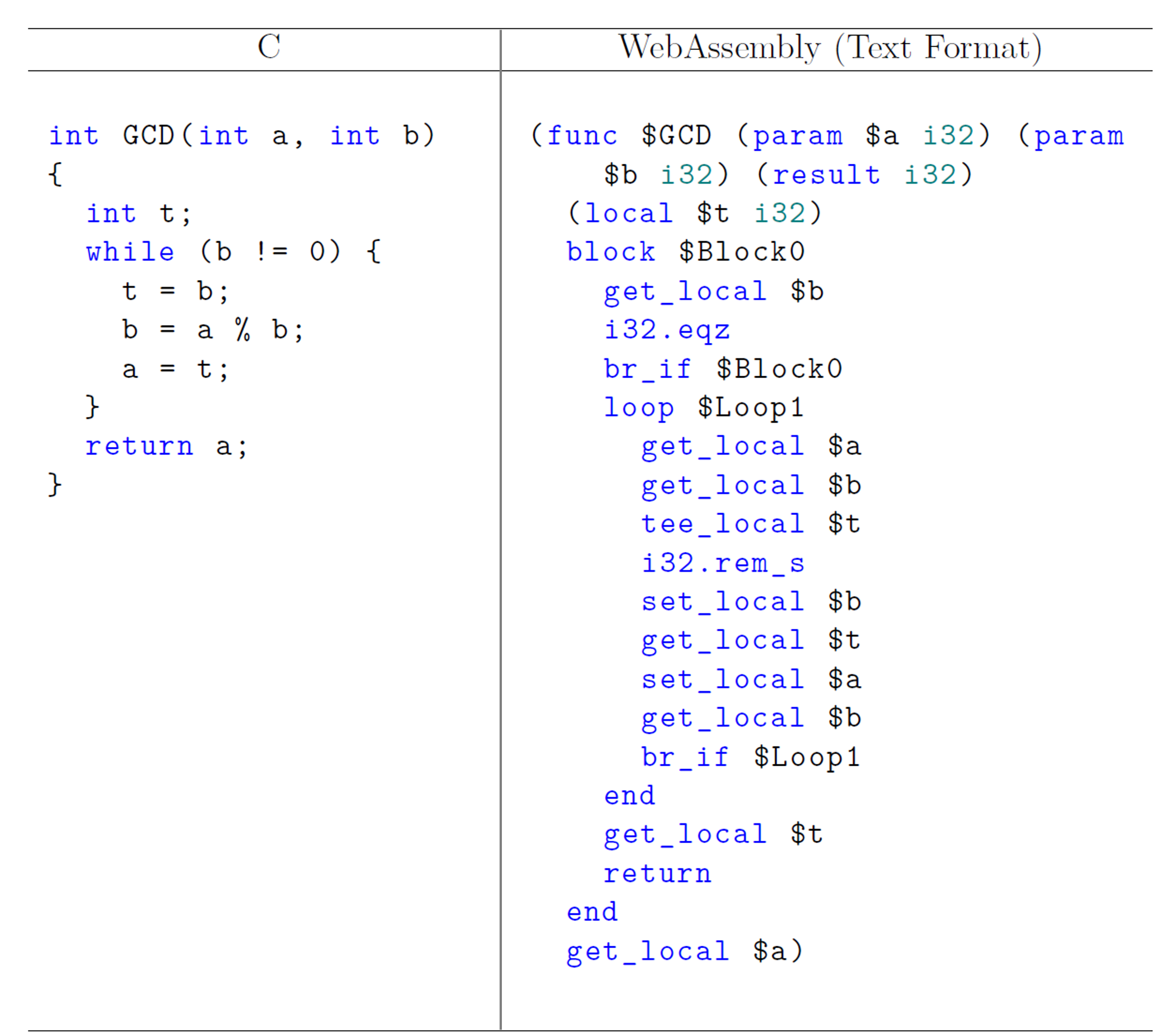}
	\caption{Euclid’s GCD algorithm implemented in C and its compiled WebAssembly code.}
	\label{fig:code1}
\end{figure}

WebAssembly is a low-level bytecode instruction set designed to be just-in-time compiled into native machine code by its host environment, typically a web browser. It serves as a compilation target for code written in performance-critical languages like C, C++, and Rust, which is then executed on a stack-based virtual machine. Due to its low-level nature and the optimizations enabled by ahead-of-time compilation, WebAssembly can achieve near-native performance while being more space-efficient compared to traditional JavaScript. Figure \ref{fig:code1} illustrates the representation of a WebAssembly module in its human-readable WebAssembly Text (WAT) format.
\subsection{WebAssembly Compilation}
Emscripten is one of the most widely adopted toolchains for compiling code into WebAssembly binaries \cite{Emscripten}. Leveraging the LLVM infrastructure \cite{LLVM:CGO04}, Emscripten enables the compilation of low-level languages such as C and C++ into WebAssembly. Additionally, Emscripten provides comprehensive support for porting legacy software to the web by offering APIs that translate \textit{OpenGL} to \textit{WebGL} and facilitate integration with popular libraries like SDL \cite{SDL}, POSIX, and pthreads \cite{pthreads}. Furthermore, Emscripten’s compatibility with Web APIs and JavaScript allows developers to create performance-optimized web applications with minimal code changes.

\subsection{WebAssembly Performance}
Over the years, several technologies have emerged with the goal of enhancing the performance of web-based applications. These include Microsoft’s \textit{ActiveX}, Adobe’s \textit{Flash platform}, and Google’s \textit{Native Client}. However, each of these technologies has since been deprecated. Another precursor to WebAssembly is \textit{asm.js} \cite{asmjs}, a subset of JavaScript aimed at optimizing performance for low-level languages. Despite its innovations, WebAssembly outperforms asm.js by supporting essential features for performance-sensitive applications, such as 64-bit integer operations, threads, and shared memory \cite{Zakai2017}.

\subsection{Browser Fingerprinting}
This section outlines our approach to developing a browser fingerprinting method that leverages variations in browser implementations to uniquely identify different web clients

\begin{figure}[tbh]
	\centering
	\includegraphics[width=0.8\linewidth]{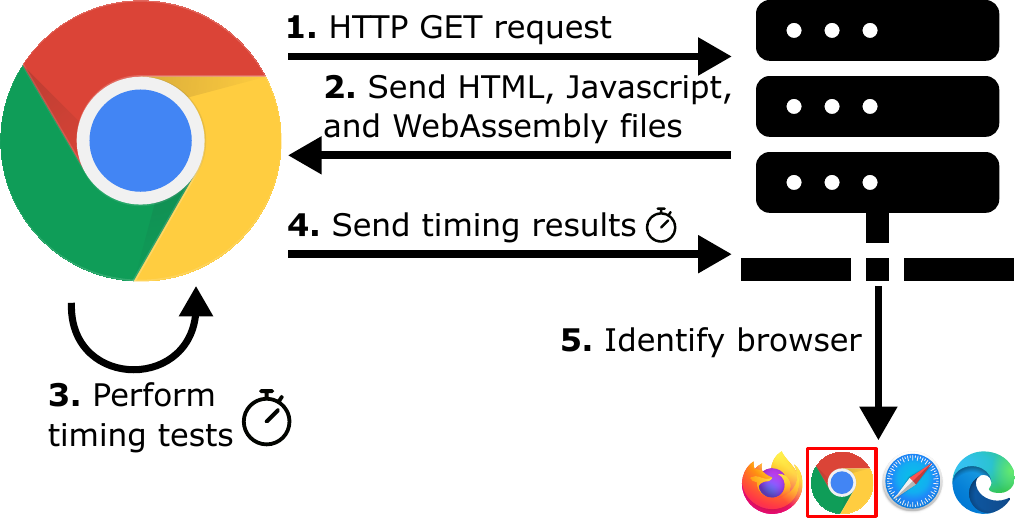}
	\caption{The proposed browser fingerprinting process}
	\label{fig:BrowserFingerprintingProcess}
\end{figure}

Figure \ref{fig:BrowserFingerprintingProcess} illustrates the complete workflow of our browser fingerprinting methodology, which leverages timing discrepancies between JavaScript and WebAssembly operations to uniquely identify web clients. The process begins when the client issues an HTTP GET request to obtain an HTML document from the web server. Upon receiving this request, the server responds by delivering the requested page, which contains embedded links to several JavaScript files that initiate the fingerprinting procedure. Subsequently, the client automatically requests these JavaScript files, which are then executed within the browser environment.

Once the JavaScript files are executed, the fingerprinting process is activated, initiating a series of timing tests that measure the performance of different interactions between JavaScript and WebAssembly. These tests are meticulously designed to capture subtle variations in execution times that arise due to differences in the underlying browser engines, hardware configurations, and operating systems. For each test, the system measures the time it takes for specific JavaScript and WebAssembly operations to complete and compares them to predefined benchmarks.

After all tests are executed, the results are aggregated into a feature vector, where each element corresponds to the timing result of a specific test. This feature vector, often referred to as the ``fingerprint," serves as a unique identifier for the client’s browser environment. The client then sends this fingerprint back to the server, where it is further analyzed and compared against a database of known fingerprints to determine the browser type and potentially identify the specific browser version or configuration.

\subsection{Technical Considerations}
In our browser fingerprinting approach, the fingerprint is represented as a high-dimensional vector of timing results, denoted as $\mathbf{fp} \in \mathbb{R}^N$, where $N$ is the total number of timing tests performed. Each timing test measures the execution latency of a particular WebAssembly or JavaScript operation, such as calling built-in mathematical functions, interacting with the WebAssembly memory space, or invoking WebAssembly functions with different argument types. The precision of these measurements is crucial, as even minor differences in timing can be indicative of distinct browser behaviors.

To further improve the accuracy of fingerprinting, our method incorporates multiple dimensions of timing tests, including:

\begin{itemize}
	\item \textbf{Wasm-to-JS Function Calls:} Tests that measure the time taken to invoke JavaScript functions from WebAssembly. This type of test highlights how different browsers optimize the interoperability between WebAssembly and JavaScript.
	\item \textbf{Memory Access Patterns:} Tests that examine the time required to read from and write to the WebAssembly linear memory. This aspect is influenced by the browser’s memory management strategies and caching mechanisms.
	\item \textbf{Built-in Function Performance:} Tests that evaluate the performance of WebAssembly calling native JavaScript functions, such as \texttt{Math.sin} or \texttt{Math.cos}. The execution times can vary based on how browsers implement these standard functions internally.
%	\item \textbf{Concurrency and Synchronization:} Tests that involve multi-threading (if supported) and the use of shared memory. These tests are designed to capture variations in the handling of concurrent operations across different browser engines.
\end{itemize}

By leveraging these diverse timing tests, our method is capable of generating a robust and unique fingerprint for each browser, capturing a wide range of performance characteristics that are difficult to replicate or obfuscate.

The feature vector generated from the tests is then compared against a pre-constructed database of known fingerprints using similarity measures such as Euclidean distance, cosine similarity, or more advanced techniques like the Mahalanobis distance. These comparisons allow the system to distinguish between different browser versions, operating systems, or even specific configurations such as installed extensions or enabled experimental features. Ultimately, this multi-dimensional approach to fingerprinting provides a detailed and comprehensive profile of the client’s browser environment, enabling not only accurate identification but also the potential for detecting subtle changes over time. For instance, if a user updates their browser or modifies specific settings, the resulting fingerprint will exhibit measurable deviations, which can be used to detect these alterations with high precision.

\begin{algorithm}[tbh]
	\hspace*{\algorithmicindent} \textbf{Input:} $timingTests$, a vector of $N$ timing tests.\\
	\begin{algorithmic}
		\STATE $fp \gets float[]$ of size $N$
		\FOR{$i \gets 1$ to $N$}
		\STATE $startTime \gets GetCurrentTime()$
		\STATE $timingTests[i]()$
		\STATE $endTime \gets GetCurrentTime()$
		\STATE $fp[i] \gets endTime - startTime$   
		\ENDFOR
		\RETURN $fp$
	\end{algorithmic}
	\caption{Algorithm for browser fingerprint generation}
	\label{alg:BrowserFPGeneration}
\end{algorithm}

Algorithm \ref{alg:BrowserFPGeneration} describes the fingerprint generation process used by the web browser. The browser performs each timing test one by one, and stores the timing results in a vector, $fp$. The vector is then sent back to the server for further analysis.

\begin{table*}[tbh]
	\centering
	\caption{List of WebAssembly Timing Tests by Mozilla}
	\label{table:BrowserTimingTests}
	\begin{tabular}{cll}
		\toprule
		\textbf{\#} & \textbf{Test Name} & \textbf{Description} \\
		\midrule
		1  & \textbf{math-builtin}       & Wasm calls into the JavaScript `Math.cos` builtin in a loop. \\
		2  & \textbf{wasm-to-js}         & Wasm calls into a JavaScript function that expects 2 arguments in a loop. \\
		3  & \textbf{call-known-0}       & Calls a monomorphic function that expects 0 arguments with 0 arguments. \\
		4  & \textbf{call-known-1}       & Calls a monomorphic function that expects 1 argument with 1 argument. \\
		5  & \textbf{call-known-2}       & Calls a monomorphic function that expects 2 arguments with 2 arguments. \\
		6  & \textbf{call-known-2-r}     & Calls a monomorphic function that expects 2 arguments with 1 argument. \\
		7  & \textbf{call-generic-2}     & Alternates between a JavaScript function call and a Wasm function, with 2 arguments (both expect 2 arguments). \\
		8  & \textbf{call-generic-2-r}   & Alternates between a JavaScript function call and a Wasm function, with 1 argument (both expect 2 arguments). \\
		9  & \textbf{scripted-getter-0}  & Calls a scripted getter that’s a Wasm function expecting 0 arguments. \\
		10 & \textbf{scripted-getter-1}  & Calls a scripted getter that’s a Wasm function expecting 1 argument. \\
		11 & \textbf{scripted-setter-1}  & Calls a scripted setter that’s a Wasm function expecting 1 argument. \\
		12 & \textbf{scripted-setter-2}  & Calls a scripted setter that’s a Wasm function expecting 2 arguments. \\
		13 & \textbf{F.p.apply-array}    & Calls a Wasm function with `Function.prototype.apply` and an array, with the expected number of arguments. \\
		14 & \textbf{F.p.apply-array-r}  & Calls a Wasm function with `Function.prototype.apply` and an array, with one fewer argument than expected. \\
		15 & \textbf{F.p.apply-args}     & Calls a Wasm function with `Function.prototype.apply` and the arguments object, with expected arguments. \\
		16 & \textbf{F.p.apply-args-r}   & Calls a Wasm function with `Function.prototype.apply` and the arguments object, with one fewer argument. \\
		17 & \textbf{F.p.call}           & Calls a Wasm function with `Function.prototype.call` and the expected number of arguments. \\
		18 & \textbf{F.p.call-r}         & Calls a Wasm function with `Function.prototype.call` and one fewer argument than expected. \\
		19 & \textbf{if-add-wasm}        & Calls a Wasm function that does: \textit{if (arg\_i32+1 != 0) return arg\_i32 + other\_arg\_i32;} \\
		20 & \textbf{if-add-js}          & Calls a JavaScript function that shouldn’t get inlined and does: \textit{if (a+1) return (a|0) + (b|0)|0;} \\
		\bottomrule
	\end{tabular}
\end{table*}

\lstdefinelanguage{WAT}{
	keywords={module, import, func, param, result, local, export, call, i32, f64, set_local, get_local, br_if, loop, i32.const, f64.const, i32.add, i32.lt_s},
	keywordstyle=\color{blue}\bfseries,
	morekeywords={loop},
	ndkeywords={},
	identifierstyle=\color{violet},
	sensitive=true,
	comment=[l]{;;},
	commentstyle=\color{gray}\ttfamily,
	morestring=[b]"
}

\lstset{
	language=WAT,
	basicstyle=\ttfamily\small,
	keywordstyle=\bfseries\color{blue},
	commentstyle=\color{gray},
	stringstyle=\color{orange},
	showstringspaces=false,
	breaklines=true,
	frame=single,
	backgroundcolor=\color{white},
	captionpos=b,
	numbers=left,
	numberstyle=\tiny\color{gray},
	xleftmargin=3em,
	framexleftmargin=2em,
	tabsize=4
}

\begin{lstlisting}[caption={WebAssembly function that calls `Math.cos` in a loop (math-builtin test \#1)}, label={lst:wasm_example}, language=WAT]
	(module
	(import "js" "cos" (func $cos (param f64) (result f64)))	
	(func $callCosInLoop (param $iterations i32) (param $angle f64) (result f64)
	(local $i i32) 
	(local $result f64) 
	(set_local $i (i32.const 0)) 
	(set_local $result (f64.const 0)) 
	(loop $loop
	(set_local $result (call $cos (get_local $angle)))
	(set_local $i (i32.add (get_local $i) (i32.const 1)))
	(br_if $loop (i32.lt_s (get_local $i) (get_local $iterations))))
	(get_local $result))
	(export "callCosInLoop" (func $callCosInLoop)))
\end{lstlisting}

Table \ref{table:BrowserTimingTests} presents the set of 20 timing tests. These timing tests cover different interactions between JavaScript and WebAssembly. We chose to focus on the four most popular web browsers: Google Chrome, Mozilla Firefox, Microsoft Chromium Edge, and Apple's Safari \cite{WebAssemblyMajorBrowsers}. The first test WebAssembly function that calls `Math.cos` in a loop (math-builtin) is shown in Listing 1.

To identify the column vector in the matrix \( A \in \mathbb{R}^{20 \times 158} \) that is most similar to a given vector \( \mathbf{b} \in \mathbb{R}^{20} \), we define the optimization problem as follows:

\[
i^* = \arg\min_{i \in \{1, 2, \ldots, 158\}} \| \mathbf{b} - \mathbf{a}_i \|_2,
\]

where \( \| \mathbf{b} - \mathbf{a}_i \|_2 \) is the Euclidean distance between the vector \( \mathbf{b} \) and the \( i \)-th column \( \mathbf{a}_i \) of matrix \( A \). This distance metric is expanded as:

\[
\| \mathbf{b} - \mathbf{a}_i \|_2^2 = \sum_{j=1}^{20} (b_j - a_{j,i})^2.
\]

This optimization problem can be expressed using inner products as follows:

\[
i^* = \arg\min_{i} \left( - 2 \mathbf{b}^\top \mathbf{a}_i + \mathbf{a}_i^\top \mathbf{a}_i \right).
\]

Moreover, incorporating a more sophisticated similarity measure such as the Mahalanobis distance allows for accounting for the variance and correlations between timing tests:

\[
i^* = \arg\min_{i} \sqrt{ (\mathbf{b} - \mathbf{a}_i)^\top \Sigma^{-1} (\mathbf{b} - \mathbf{a}_i) },
\]

where \( \Sigma \) represents the covariance matrix of the test results, capturing the interdependencies among features.

To further enhance the accuracy of the similarity comparison, fingerprinting can also employ Principal Component Analysis (PCA). By projecting each column \( \mathbf{a}_i \) of \( A \) onto a reduced-dimensional subspace defined by the principal components, the transformed matrix \( A' \in \mathbb{R}^{k \times 158} \) is obtained, where \( k \) is the number of principal components retained. Similarly, projecting \( \mathbf{b} \) onto this subspace, denoted as \( \mathbf{b}' \), allows us to redefine the optimization problem as:

\[
i^* = \arg\min_{i} \| \mathbf{b}' - \mathbf{a}'_i \|_2,
\]

where \( \| \mathbf{b}' - \mathbf{a}'_i \|_2 \) is the Euclidean distance in the reduced-dimensional subspace. This approach reduces noise and dimensionality, leading to a more robust comparison between vectors. The PCA-based technique leverages the variance captured by the principal components, thereby improving the effectiveness of the fingerprinting comparison and identification process.

\section{Evaluation}
The evaluation of our browser fingerprinting methodology was conducted on a diverse set of devices and configurations to ensure robustness and generalizability. Specifically, we focused on assessing the implementation of JavaScript and WebAssembly interactions across multiple browsers and operating systems. The hardware and software configurations used for the evaluation are detailed in Table \ref{table:BrowserFingerprintingDevices}.

A total of 25 physical devices were employed for testing, including 14 PC workstations, a MacBook Pro, six Android devices, and four iOS devices. The operating systems spanned six different platforms: Windows, macOS, CentOS, Ubuntu, Android, and iOS. The Android devices consisted of three Google Pixel devices, two OnePlus devices, and one Samsung Galaxy S20. Meanwhile, the iOS devices included three iPhones and one iPad. Additionally, virtual machines were utilized to simulate environments on four hypervisors: VMWare, Hyper-V, VirtualBox, and KVM. These configurations enabled us to perform fingerprinting on a range of environments, each representing different performance and interaction characteristics.

\begin{table}[]
	\centering
	\begin{tabular}{|l|l|c|c|l|}
		\hline
		\textbf{Device} & \textbf{CPU} & \textbf{Cores} & \textbf{RAM (GB)} & \textbf{OS} \\ \hline
		PC1  & Intel i7-4770K  & 8 & 32 & Windows \\ 
		PC2  & Intel i7-9700K  & 4 & 6  & Windows \\ 
		PC3  & Intel i7-9700K  & 4 & 16 & Windows \\ 
		PC4  & Intel i5-4690K  & 8 & 32 & Windows \\ 
		PC5  & Intel i7-8700   & 8 & 4  & Windows \\ 
		PC6  & Intel i7-7700HQ & 8 & 32 & Windows \\ 
		PC7  & AMD Ryzen 9 5900X & 8 & 4 & Windows \\ 
		PC8  & AMD Athlon 3000G & 4 & 12 & Windows \\ 
		PC9  & Intel i7-9700K  & 6 & 6  & Windows \\ 
		PC10 & Intel i5-4690   & 4 & 3  & Windows \\ 
		PC11 & Intel i5-3470   & 4 & 2  & Windows \\ 
		PC12 & Intel i7-4790   & 4 & 16 & Windows \\ 
		PC13 & AMD Ryzen 9 5900X & 4 & 8  & Windows \\ 
		PC14 & AMD Ryzen 9 3900X & 16 & 32 & Windows \\ 
		Mac1 & Intel i5-8257U  & 4 & 16 & macOS \\ 
		IOS1 & Hexa-core       & 8 & 4  & iOS \\ 
		IOS2 & Hexa-core       & 4 & 3  & iOS \\ 
		IOS3 & Hexa-core       & 8 & 16 & iPadOS \\ 
		IOS4 & Hexa-core       & 8 & 8  & iOS \\ 
		Android1 & Octa-core  & 4 & 8  & Android \\ 
		Android2 & Octa-core  & 4 & 8  & Android \\ 
		Android3 & Octa-core  & 8 & 16 & Android \\ 
		Android4 & Octa-core  & 8 & 16 & Android \\ 
		Android5 & Quad-core  & 12 & 32 & Android \\ 
		Android6 & Octa-core  & 4 & 6  & Android \\ \hline 
	\end{tabular}
	\caption{Device specifications used for browser fingerprinting evaluation.}
	\label{table:BrowserFingerprintingDevices}
\end{table}

%Figure \ref{fig:BrowserFingerprintingDevices} depicts the distribution of operating systems, hypervisors, and web browsers utilized across the different device types. For Windows, the hypervisors tested included VMWare, Hyper-V, and VirtualBox. In contrast, for CentOS and Ubuntu, VMWare, VirtualBox, and KVM were utilized. We created three virtual machines per operating system and hypervisor combination, which enabled us to capture variations in behavior across different virtualized environments.

\begin{figure}[]
	\centering
	\includegraphics[width=1\linewidth]{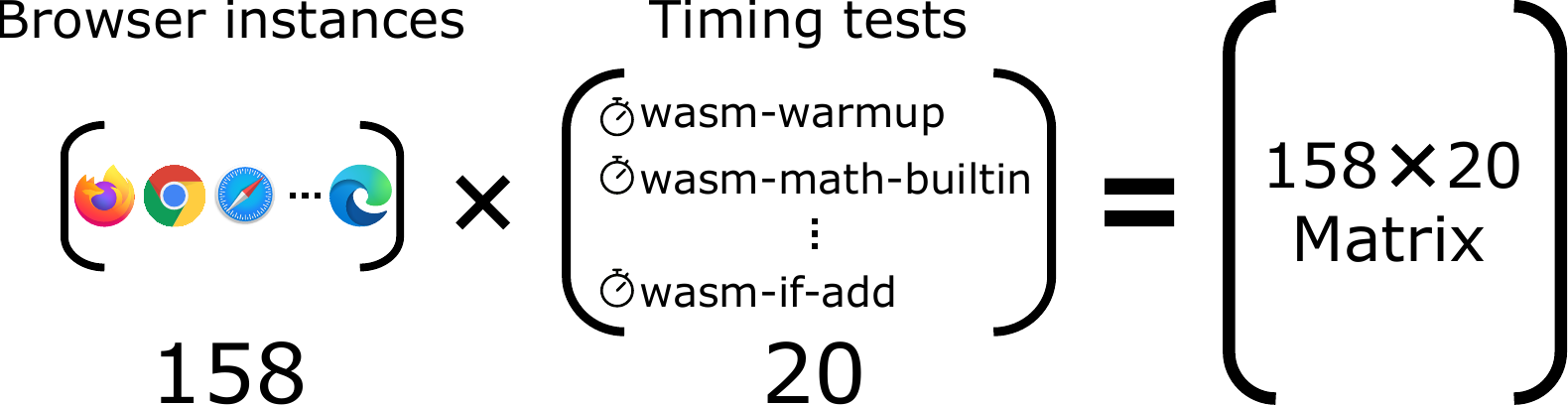}
	\caption{Overview of 158 browser instances, each running 20 timing tests.}
	\label{fig:BrowserInstancesTimingTests}
\end{figure}

The web browsers examined in this study were Google Chrome, Microsoft Chromium Edge, and Mozilla Firefox for the Windows, CentOS, Ubuntu, and Android platforms. For macOS and iOS, the browsers included Google Chrome, Microsoft Chromium Edge, Mozilla Firefox, and Apple’s Safari. Figure \ref{fig:BrowserInstancesTimingTests} provides an overview of the 158 unique browser instances used for this evaluation. Each browser instance was subjected to 20 timing tests as detailed in Table \ref{table:BrowserTimingTests}, generating comprehensive data on their interactions.

The WebAssembly timing tests were conducted across bare-metal devices, virtualized environments, and smartphones. The results highlight significant performance variations depending on the operating system, browser, and device configurations.

\subsubsection{Bare-Metal Windows (Chrome, Edge, Firefox)} 
\begin{figure}[]
	\centering
	\includegraphics[width=1.0\linewidth]{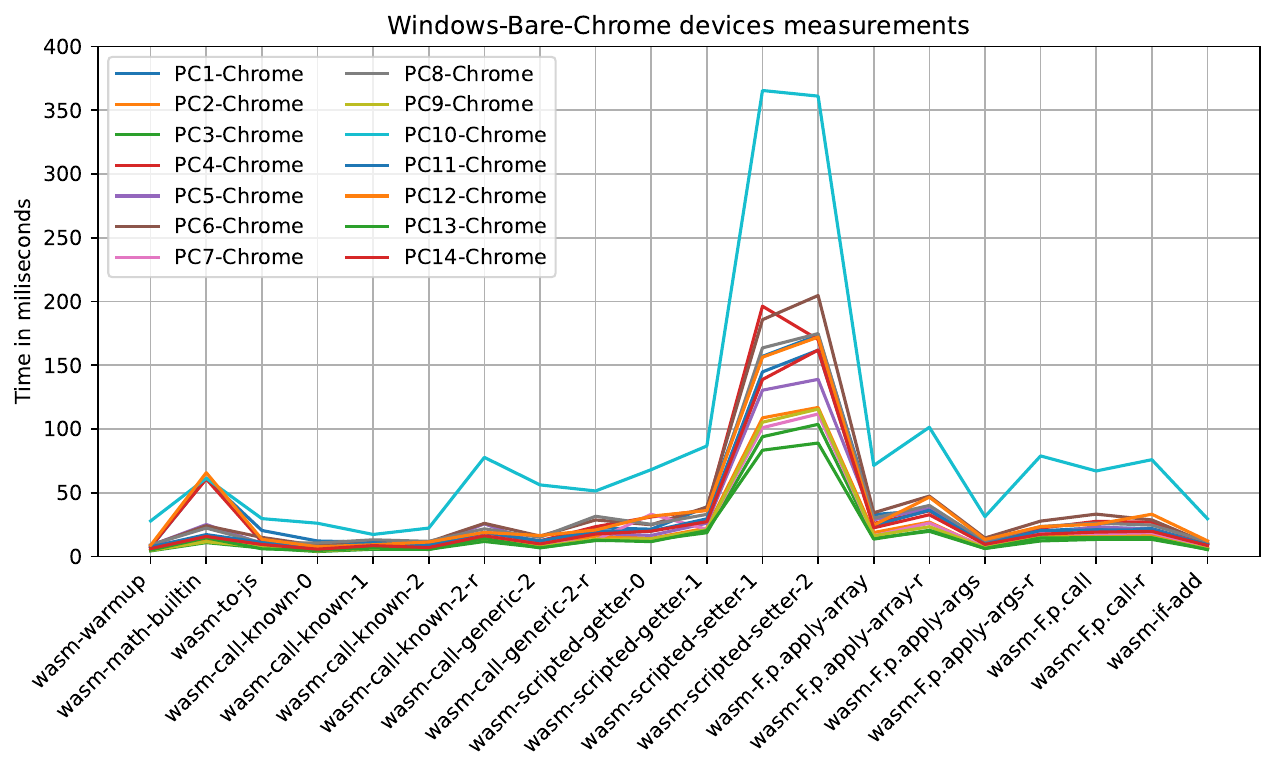}
	\caption{Timing results of WebAssembly tests on bare-metal Windows devices using Chrome.}
	\label{fig:graphBrowserBareWindowsChrome}
\end{figure}

Figure \ref{fig:graphBrowserBareWindowsChrome} shows the timing results for Chrome on bare-metal Windows devices. The mean timing results for `wasm-scripted-setter-1` and `wasm-scripted-setter-2` are significantly higher (152.21ms and 161.14ms, respectively) compared to other tests (19.54ms). Similar trends are observed for Edge and Firefox, indicating that `wasm-scripted-setter` tests are more sensitive to system variations.

\subsubsection{Bare-Metal Unix (CentOS, Ubuntu, macOS)} 
\begin{figure}[]
	\centering
	\includegraphics[width=1.1\linewidth]{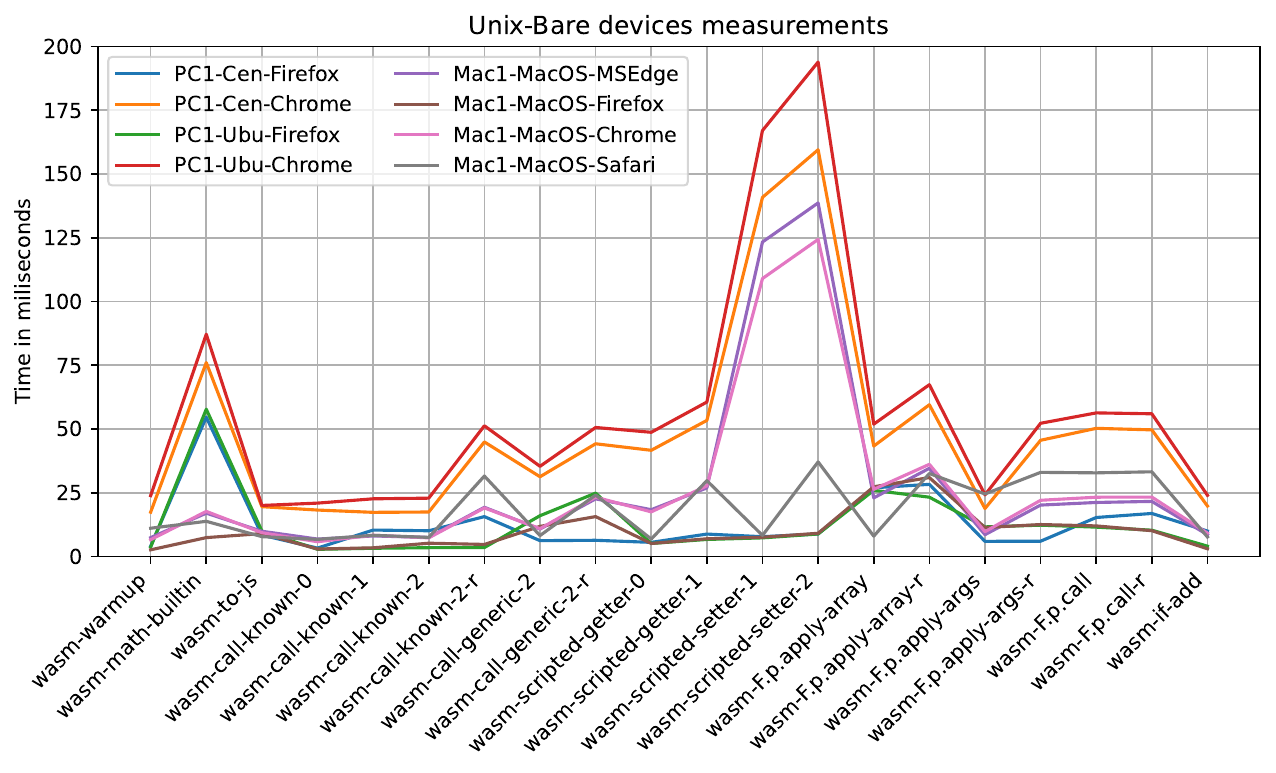}
	\caption{Timing results of WebAssembly tests on bare-metal Unix devices.}
	\label{fig:graphBrowserBareUnix}
\end{figure}

Figure \ref{fig:graphBrowserBareUnix} presents the timing results for Unix-like operating systems. For Chrome, the `wasm-scripted-setter` tests show 300-750\% increased mean times compared to other tests. Firefox and Safari exhibit more consistent results with lower deviations.

\subsubsection{Virtualized Environments (Hyper-V, VMWare, VirtualBox, KVM)} 
\begin{figure}[]
	\centering
	\includegraphics[width=1\linewidth]{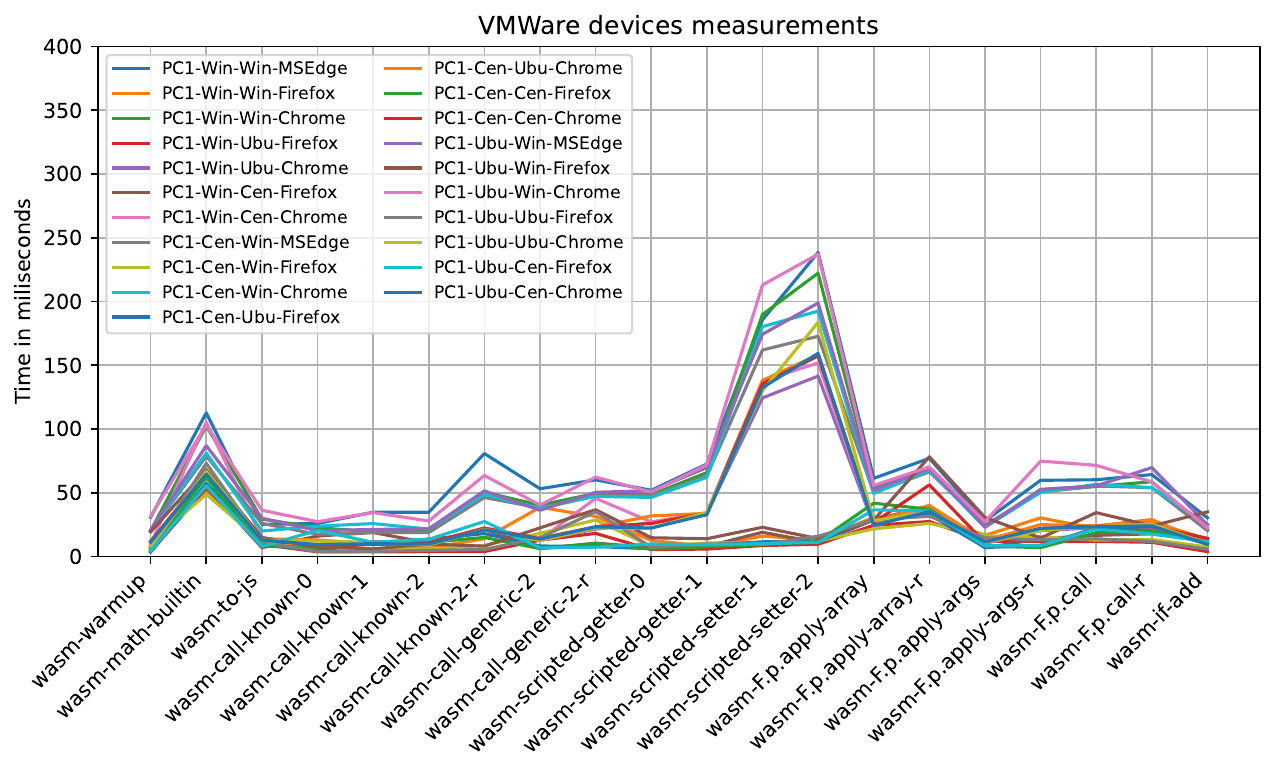}
	\caption{Timing results of WebAssembly tests for VMs running on VMWare hypervisor.}
	\label{fig:graphBrowserVMWare}
\end{figure}

The timing results for VMs running on different hypervisors show Chrome and Edge exhibiting up to 570\% increased times for `wasm-scripted-setter` tests compared to others. Firefox maintains consistent results across all hypervisors. Figure \ref{fig:graphBrowserVMWare} show the VMWare measurements.

\subsubsection{Smartphones (Android, iOS)} 
\begin{figure}[]
	\centering
	\includegraphics[width=1.0\linewidth]{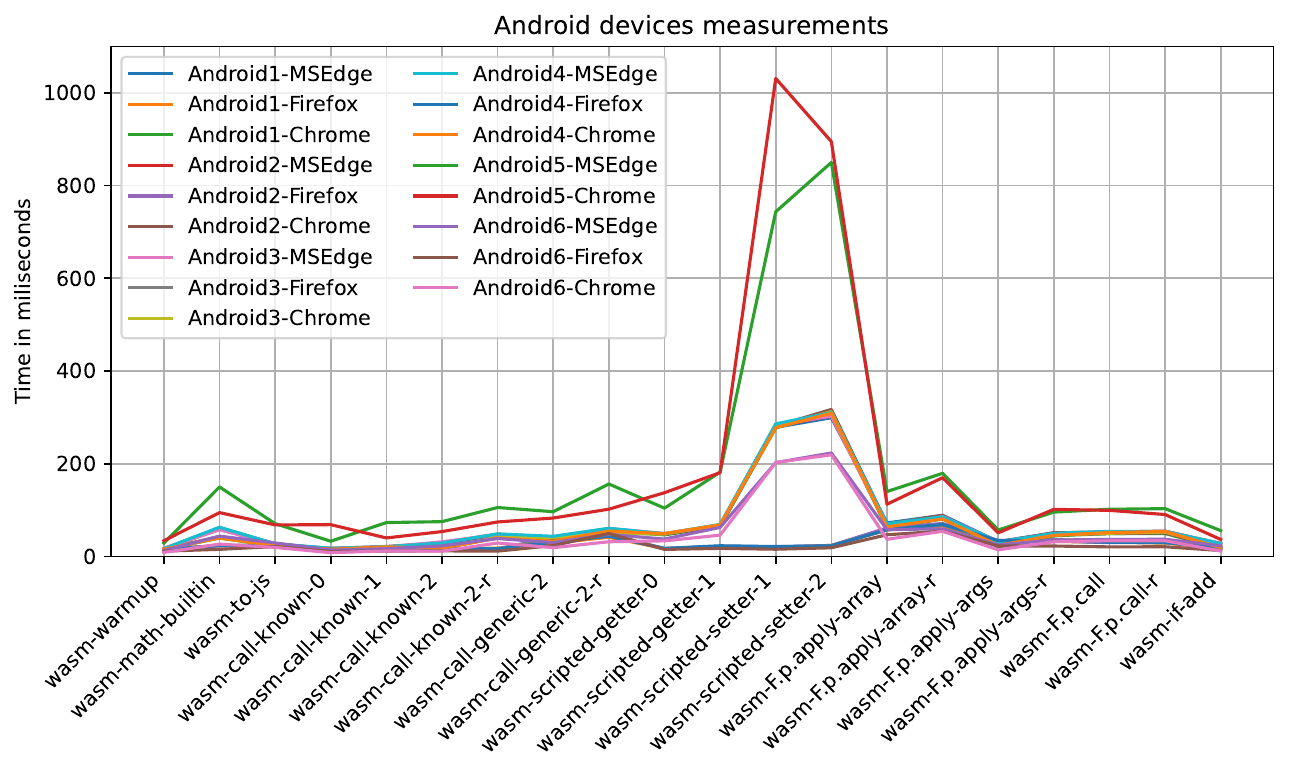}
	\caption{Timing results of WebAssembly tests for Android smartphones.}
	\label{fig:graphBrowserAndroid}
\end{figure}

\begin{figure}[]
	\centering
	\includegraphics[width=1.0\linewidth]{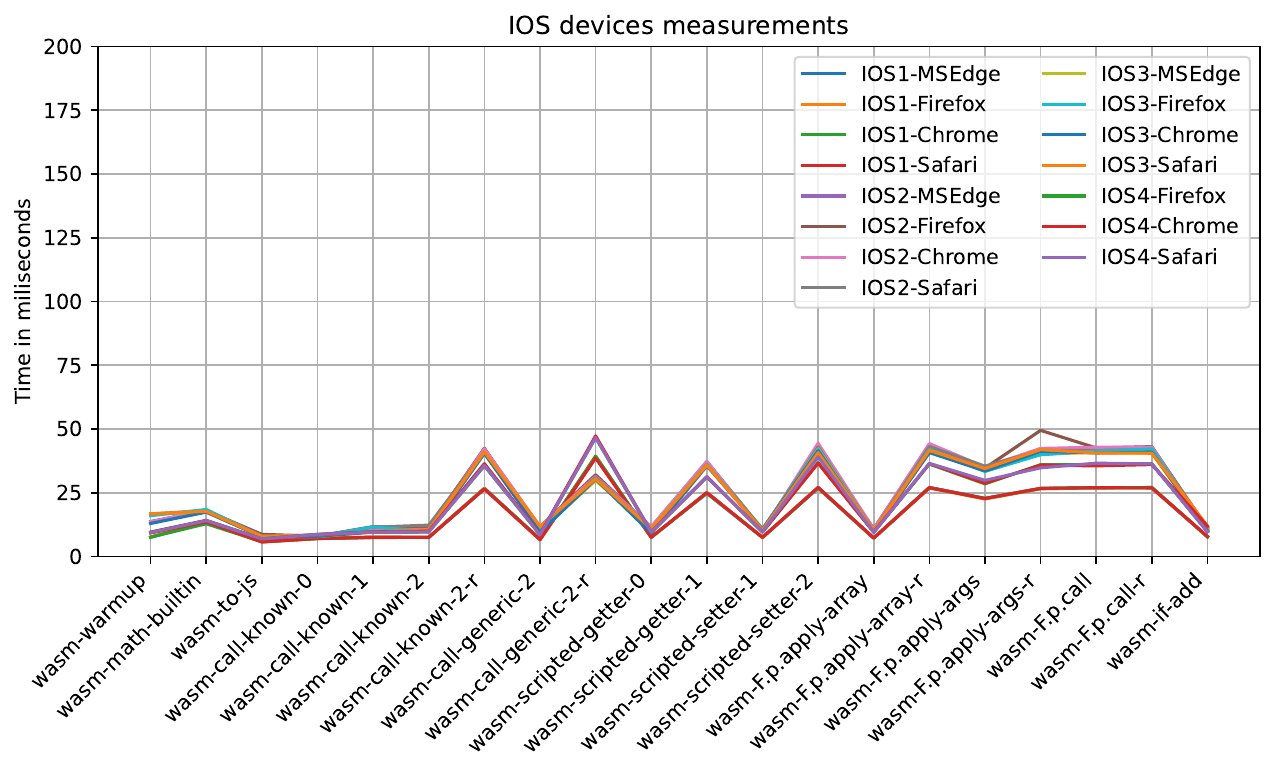}
	\caption{Timing results of WebAssembly tests for iOS smartphones.}
	\label{fig:graphBrowserIOS}
\end{figure}

The timing results for Android (Figure \ref{fig:graphBrowserAndroid}) show high variability, with `wasm-scripted-setter` tests 700-760\% slower than others. iOS devices (Figure \ref{fig:graphBrowserIOS}) have more consistent results across all browsers with mean times around 21ms.

The findings indicate that `wasm-scripted-setter` tests are particularly influenced by system and browser variations, making them valuable for browser fingerprinting across different environments.

\subsubsection{Chromium Based Browser Identification}
The timing test results revealed that on all device types, excluding iOS, there are substantial differences in the execution times of \textit{wasm-scripted-setter-1} and \textit{wasm-scripted-setter-2} when compared to the other timing tests in both Chrome and Edge browsers. This disparity was not observed when using Firefox and Safari, indicating that the timing characteristics of these tests can be used to distinguish between Chromium-based and non-Chromium-based browsers.

\lstset{
	language=JavaScript,
	basicstyle=\ttfamily\small,
	keywordstyle=\color{blue},
	stringstyle=\color{red},
	commentstyle=\color{green},
	numbers=left,
	numberstyle=\tiny,
	stepnumber=1,
	numbersep=0pt,
	showstringspaces=false,
	tabsize=1,
	breaklines=true,
	breakatwhitespace=false,
}

\begin{flushleft}
\begin{algorithm}[]
	\caption{\textit{wasm-scripted-setter-1} and \textit{wasm-scripted-setter-2} Implementation}
	\label{alg:Setter1Setter2}
	\vspace{5pt}
	\begin{lstlisting}[language=JavaScript]
		var exports = WebAssemblyInstance.exports;
		// JavaScript implementation of wasm-scripted-setter-1
		function wasm_scripted_setter_1(limit) {
			let GETSET = {};
			Object.defineProperty(GETSET, 'x', {
				set: exports.set_global_one  // Define setter using WebAssembly function
			});
			for (var i = 0; i < limit; i++) {
				GETSET.x = i;  // Trigger the setter with incremental values
			}
		}	
		// JavaScript implementation of wasm-scripted-setter-2
		function wasm_scripted_setter_2(limit) {
			let GETSET = {};
			Object.defineProperty(GETSET, 'x', {
				set: exports.set_global_two  // Define setter with a WebAssembly function
			});
			for (var i = 0; i < limit; i++) {
				GETSET.x = i;  // Trigger the setter with incremental values
			}
		}
	\end{lstlisting}
\end{algorithm}
\end{flushleft}

Algorithm \autoref{alg:Setter1Setter2} illustrates the JavaScript implementation of \textit{wasm-scripted-setter-1} and \textit{wasm-scripted-setter-2}. Both tests utilize \textit{Object.defineProperty} to define a setter function for the property \texttt{"x"} of the \texttt{GETSET} object. In these timing tests, the \texttt{exports} object represents the WebAssembly module exports. In \textit{wasm-scripted-setter-1}, the setter function is \texttt{exports.set\_global\_one}, which is a WebAssembly function that takes a single argument. Similarly, in \textit{wasm-scripted-setter-2}, the setter function is \texttt{exports.set\_global\_two}, which requires two arguments.

The variation in timing test results between Chromium-based browsers and non-Chromium browsers can be exploited by comparing the results of \textit{wasm-scripted-setter-1} and \textit{wasm-scripted-setter-2} against a baseline, such as \textit{wasm-scripted-getter-0}. The following algorithm uses these comparisons to classify whether the browser is Chromium-based or not.

\begin{algorithm}
	\caption{$\text{isChromium}(SS1, SS2, SG0, SS1Threshold, SS2Threshold)$: Determine if Browser is Chromium-Based}
	\label{alg:isChromium}
	
	\begin{algorithmic}
		\STATE \textbf{Inputs:}
		\STATE $SS1$: Timing result of \textit{wasm-scripted-setter-1}
		\STATE $SS2$: Timing result of \textit{wasm-scripted-setter-2}
		\STATE $SG0$: Timing result of \textit{wasm-scripted-getter-0}
		\STATE $SS1Threshold$: Threshold value for the ratio $SS1/SG0$
		\STATE $SS2Threshold$: Threshold value for the ratio $SS2/SG0$

		\STATE \textbf{Return:}
		\STATE ($SS1/SG0 \geq SS1Threshold$) AND ($SS2/SG0 \geq SS2Threshold$)
	\end{algorithmic}
\end{algorithm}

Algorithm \autoref{alg:isChromium} presents the method for determining if a browser is Chromium-based. The algorithm uses the ratios of \textit{wasm-scripted-setter-1} and \textit{wasm-scripted-setter-2} relative to \textit{wasm-scripted-getter-0} and compares them against predefined threshold values. By analyzing these ratios, the algorithm can accurately classify the browser type.

\begin{table}[]
	\centering
	\caption{Descriptive statistics of the ratios between \textit{wasm-scripted-setter-[1,2]} and \textit{wasm-scripted-getter-0} in Chrome, Edge, and Firefox.}
	\label{table:BrowserChromiumStats}
	\resizebox{\columnwidth}{!}{ % Resize the table to fit within a single column
		\begin{tabular}{lccc|ccc}
			\toprule
			& \multicolumn{3}{c}{\textbf{ScriptedSetter1/ScriptedGetter0}} & \multicolumn{3}{c}{\textbf{ScriptedSetter2/ScriptedGetter0}} \\
			\textbf{Statistic} & \textbf{All} & \textbf{Firefox} & \textbf{Chrome + Edge} & \textbf{All} & \textbf{Firefox} & \textbf{Chrome + Edge} \\
			\midrule
			\textbf{Count} & 142 & 55 & 87 & 142 & 55 & 87 \\
			\textbf{Mean} & 4.07 & 1.67 & 5.59 & 4.57 & 1.98 & 6.21 \\
			\textbf{Std} & 2.19 & 0.68 & 1.26 & 2.54 & 1.57 & 1.40 \\
			\cmidrule(lr){1-7}
			\textbf{Min} & 0.51 & 0.51 & 3.05 & 0.62 & 0.62 & 3.38 \\
			\textbf{5\%} & 1.09 & 0.93 & 3.41 & 1.12 & 0.91 & 3.84 \\
			\textbf{25\%} & 1.74 & 1.23 & 4.46 & 1.96 & 1.35 & 5.09 \\
			\cmidrule(lr){1-7}
			\textbf{50\%} & 4.24 & 1.56 & 5.77 & 4.71 & 1.78 & 6.25 \\
			\textbf{75\%} & 5.92 & 1.89 & 6.64 & 6.60 & 2.06 & 7.48 \\
			\cmidrule(lr){1-7}
			\textbf{95\%} & 7.29 & 2.82 & 7.48 & 8.14 & 3.10 & 8.20 \\
			\textbf{Max} & 7.88 & 4.31 & 7.88 & 12.30 & 12.30 & 8.40 \\
			\bottomrule
		\end{tabular}
	}
\end{table}

\autoref{table:BrowserChromiumStats} presents the descriptive statistics of the ratios between \textit{wasm-scripted-setter-1} and \textit{wasm-scripted-setter-2} relative to \textit{wasm-scripted-getter-0} for Chrome, Edge, and Firefox. From these statistics, we set the threshold values at $SS1Threshold = 3.05$ and $SS2Threshold = 3.10$. Using these thresholds, all Chromium-based browsers exceeded the threshold values, while less than 5\% of Firefox instances did. This configuration resulted in a classification success rate of 99.29\%, with only one misclassified Firefox instance out of 55.

These results demonstrate the effectiveness of our method in differentiating between Chromium-based and non-Chromium-based browsers using the proposed timing tests and threshold-based classification algorithm.

\begin{figure}[]
	\centering
	\includegraphics[width=1\linewidth]{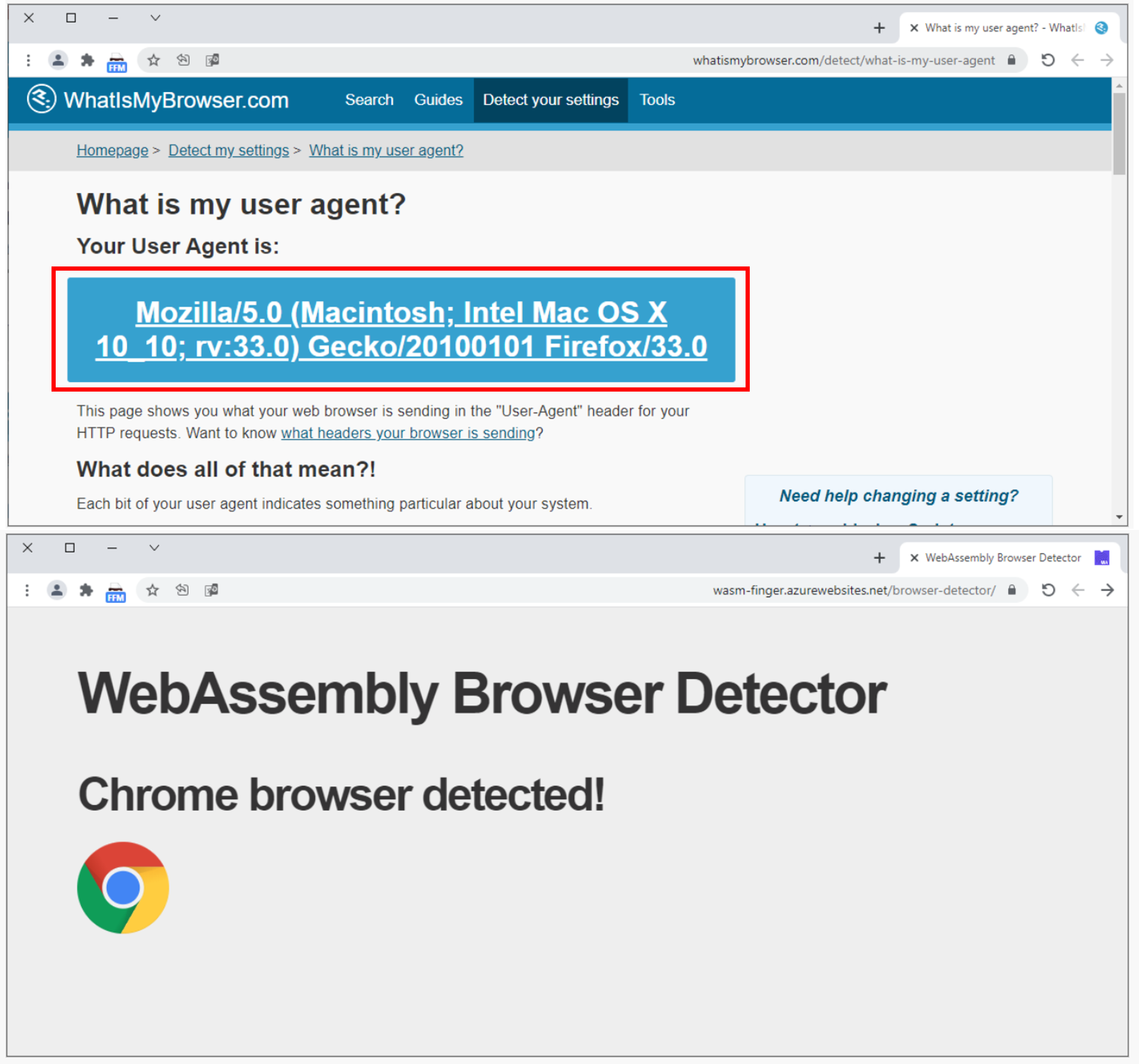}
	\caption{A browser with reported spoofed Firefox 33.0 user-agent (top) and the detected real browser (bottom)}
	\label{fig:spoof}
\end{figure}

Figure \ref{fig:spoof} shows the screenshot of a browser with reported spoofed user-agent (left) and the detected real browser (right).

Across all tested configurations:
\begin{itemize}
	\item \textbf{Bare-Metal Windows}:
	\begin{itemize}
		\item For Chrome, `wasm-scripted-setter-1` and `wasm-scripted-setter-2` were 678.96\% and 724.66\% slower than other tests, respectively.
		\item For Edge, `wasm-scripted-setter-1` and `wasm-scripted-setter-2` were 568.10\% and 651.74\% slower than other tests, respectively.
	\end{itemize}
	
	\item \textbf{Bare-Metal Unix}:
	\begin{itemize}
		\item For Chrome, `wasm-scripted-setter-1` and `wasm-scripted-setter-2` were 329.32\% and 391.97\% slower than other tests, respectively.
		\item Firefox and Safari displayed consistent results, with minimal deviation across all timing tests.
	\end{itemize}
	
	\item \textbf{Virtualized Environments (Hyper-V, VMWare, VirtualBox, KVM)}:
	\begin{itemize}
		\item On KVM, Chrome's `wasm-scripted-setter-1` and `wasm-scripted-setter-2` tests were 402.45\% and 512.67\% slower than other tests, respectively.
		\item Firefox showed consistent timing results across all hypervisors with minimal deviations.
	\end{itemize}
	
	\item \textbf{Smartphones (Android, iOS)}:
	\begin{itemize}
		\item On Android, `wasm-scripted-setter-1` and `wasm-scripted-setter-2` were up to 760.87\% slower than other tests, showing high variability.
		\item iOS devices showed stable results, with mean timings around 21ms across all browsers.
	\end{itemize}
\end{itemize}

These findings demonstrate that the `wasm-scripted-setter` tests are particularly influenced by system and browser variations, making them a valuable metric for browser fingerprinting in diverse environments.

%\vspace{15mm} % Adjust the negative space to your liking
\subsection{Mitigation}
The proposed browser fingerprinting method leverages timing discrepancies in interactions between JavaScript and WebAssembly to identify Chromium-based browsers. This technique is particularly effective due to the pronounced execution time differences in `wasm-scripted-setter-1` and `wasm-scripted-setter-2` tests, which involve setting WebAssembly functions as setters using the \texttt{Object.defineProperty} function.

To counteract this fingerprinting approach, we introduce a mitigation strategy that injects random delays into the execution of setter functions in Chromium and non-Chromium browsers, such as Firefox. By manipulating the timing results, the mitigation alters the timing ratios between tests, making Firefox appear similar to a Chromium-based browser and thereby reducing the effectiveness of the fingerprinting method. Algorithm \ref{alg:MitigationBrowserAddDelay} presents the JavaScript implementation of this approach, where the original \texttt{Object.defineProperty} method is hooked to replace the assigned setter with a delayed version.
\begin{algorithm}[]
	\caption{Inject Delay to Setters of Object Properties}
	\label{alg:MitigationBrowserAddDelay}
	\vspace{5pt} % Add space between the caption and the code
	\begin{lstlisting}[language=JavaScript]
		// Random delay for the function
		let delay = Math.random() * 1000;
		let originalDefineProperty = Object.defineProperty;
		// Inject delay into setter functions
		function myDefineProperty(obj, prop, descr) {
			if ('set' in descr && typeof descr['set'] === "function") {
				descr['set'] = createDelayedFunction(descr['set'], delay);  
			}
			originalDefineProperty(obj, prop, descr);  // Call original defineProperty
		}
	%	// Replace Object.defineProperty 
		Object.defineProperty = myDefineProperty;
	\end{lstlisting}
\end{algorithm}

\begin{figure}[]
	\centering
	\includegraphics[width=0.9\linewidth]{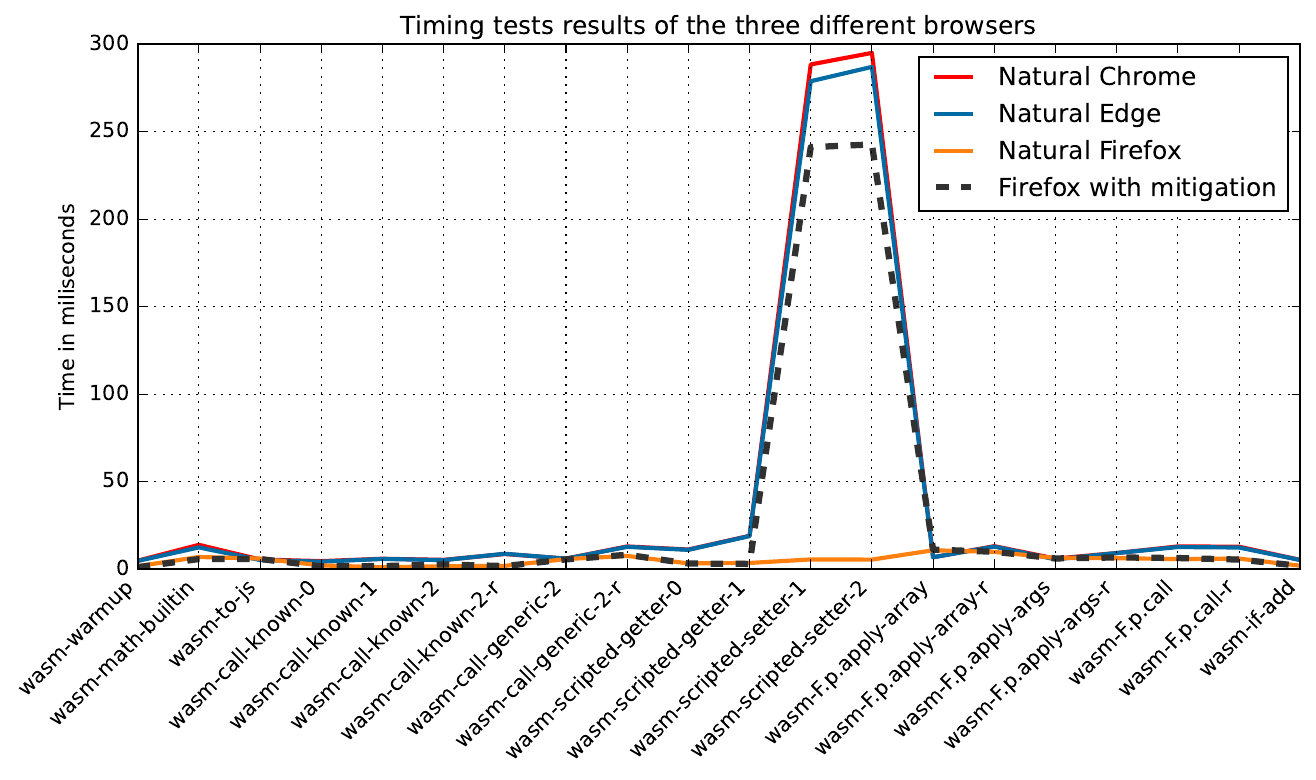}
	\caption{WebAssembly timing test results of Chrome, Edge, Firefox, and Firefox with mitigation.}
	\label{fig:mitigation-browser}
\end{figure}

Figure \ref{fig:mitigation-browser} compares the timing test results for Chrome, Edge, Firefox, and Firefox with the proposed mitigation applied. For Chrome and Edge, the timing ratios of `wasm-scripted-setter-1` and `wasm-scripted-setter-2` relative to `wasm-scripted-getter-0` were 25.42 and 26.00 (Chrome), and 24.93 and 25.66 (Edge), respectively. Without mitigation, Firefox exhibited lower ratios of 1.65 and 1.63. However, with the mitigation in place, the ratios for Firefox increased to 71.38 and 71.85, making its characteristics closer to those of Chromium-based browsers.

%\begin{figure}[H]
%	\centering
%	\includegraphics[width=1\linewidth]{fig/algmit.png}
%	\label{fig:algmit}
%\end{figure}

This mitigation strategy demonstrates that by artificially adjusting the timing patterns of specific functions, the distinguishing features used by the fingerprinting method are obscured. Consequently, non-Chromium browsers, such as Firefox, can be made to appear more like Chromium-based browsers, effectively reducing the overall fingerprinting accuracy and introducing potential false negatives.

\subsection{Implementation of a WebAssembly Hooking Module with Random Delays}

The implementation introduces a mechanism for intercepting WebAssembly (Wasm) functions and modifying their behavior by adding random delays. This process leverages the JavaScript WebAssembly API to dynamically hook into the exported functions of a compiled WebAssembly module. The first step involves loading and compiling the WebAssembly module, which is achieved using the API’s \texttt{WebAssembly.instantiate} function. This function takes a binary \texttt{.wasm} file as input and returns an instance of the module with access to its exported functions. Below is an example of how the module is loaded:

\begin{lstlisting}[language=JavaScript, caption={Loading and Compiling a WebAssembly Module}]
	const loadWasm = async (wasmPath) => {
		const response = await fetch(wasmPath);
		const bytes = await response.arrayBuffer();
		return WebAssembly.instantiate(bytes, {});
	};
\end{lstlisting}

Once the module is compiled, the implementation wraps its exports in a proxy to intercept function calls. For each exported function, a new asynchronous function is introduced. This function introduces a random delay using JavaScript’s \texttt{Promise} and \texttt{setTimeout} mechanisms, logs the delay for debugging purposes, and then invokes the original function. The random delay is generated by scaling a random value to the desired range, as shown in the following code:

\begin{lstlisting}[language=JavaScript, caption={Hooking WebAssembly Exports to Add Random Delays}]
	const hookExports = (wasmExports) => {
		const hookedExports = {};
		for (const [key, func] of Object.entries(wasmExports)) {
			if (typeof func === "function") {
				hookedExports[key] = async (...args) => {
					const delay = Math.random() * 200; // Delay up to 200 milliseconds
					console.log(`Hooked: ${key}, introducing a delay of ${delay.toFixed(2)} ms`);
					await new Promise((resolve) => setTimeout(resolve, delay));
					return func(...args); // Call the original function
				};
			} else {
				hookedExports[key] = func;
			}
		}
		return hookedExports;
	};
\end{lstlisting}

The hooked functions are then integrated into the WebAssembly instance, replacing the original exports with their delayed versions. This integration ensures that all calls to the exports pass through the delay mechanism, while still preserving the original functionality. An example of using the hooked instance is provided below:

\begin{lstlisting}[language=JavaScript, caption={Integrating Hooked Functions into a WebAssembly Instance}]
	(async () => {
		const wasmModule = await loadWasm("module.wasm");
		const hookedInstance = hookExports(wasmModule.instance.exports);
		
		// Example usage of hooked exports
		const result = await hookedInstance.someFunction(10, 20);
		console.log("Result:", result);
	})();
\end{lstlisting}

This design introduces non-blocking random delays that allow for asynchronous execution, making it suitable for testing or obfuscation scenarios. Adding random delays impacts performance, so it should be used judiciously, particularly in environments requiring high throughput. Debugging logs provide visibility into the delays and intercepted functions, aiding in verifying the behavior of the hooking mechanism. Overall, this approach demonstrates a flexible method for modifying WebAssembly functionality at runtime, leveraging JavaScript’s capabilities to extend Wasm's behavior in dynamic and non-intrusive ways.

\subsection{Detecting Timing Attacks on WebAssembly for Fingerprinting}
To counteract the FingerPrinting behavior, a detection mechanism can be implemented that monitors access to timing functions and profiles WebAssembly execution for suspicious patterns.

The detection begins by intercepting calls to high-resolution timers. This is achieved by overriding the default implementation of \texttt{performance.now()} to include logging and monitoring of access frequency. If the frequency of calls exceeds a predefined threshold, the system flags the activity as suspicious. The following code illustrates this approach:

\begin{lstlisting}[language=JavaScript, caption={Intercepting High-Resolution Timer Access}]
	const originalPerformanceNow = performance.now;
	performance.now = function () {
		const timestamp = originalPerformanceNow.apply(this, arguments);
		console.log(`High-resolution timer accessed: ${timestamp}`);
		detectSuspiciousTimingAccess();
		return timestamp;
	};
	
	let timerAccessCount = 0;
	const timerAccessThreshold = 100;
	
	function detectSuspiciousTimingAccess() {
		timerAccessCount++;
		if (timerAccessCount > timerAccessThreshold) {
			console.warn("Suspicious timing attack detected: excessive timer access");
		}
	}
\end{lstlisting}

In addition to intercepting timers, WebAssembly exports can be instrumented to monitor execution times of individual functions. Each function call is wrapped in a mechanism that measures the time taken for execution. Functions with extremely short execution times, particularly when called repeatedly, are flagged as potentially being used in timing attacks. An example of such instrumentation is shown below:

\begin{lstlisting}[language=JavaScript, caption={Profiling WebAssembly Exported Functions}]
	const hookWasmExportsForTiming = (wasmExports) => {
		const hookedExports = {};
		for (const [key, func] of Object.entries(wasmExports)) {
			if (typeof func === "function") {
				hookedExports[key] = async (...args) => {
					const startTime = performance.now();
					const result = func(...args); // Call the original function
					const endTime = performance.now();
					const executionTime = endTime - startTime;
					
					console.log(`Function ${key} executed in ${executionTime} ms`);
					if (executionTime < 1) {
						console.warn(`Suspicious timing detected on function ${key}`);
					}
					return result;
				};
			} else {
				hookedExports[key] = func;
			}
		}
		return hookedExports;
	};
	
	(async () => {
		const wasmModule = await WebAssembly.instantiateStreaming(fetch('module.wasm'), {});
		const hookedInstance = hookWasmExportsForTiming(wasmModule.instance.exports);
	})();
\end{lstlisting}

Detecting timing attacks presents several challenges. Timing discrepancies that appear suspicious in one context might be legitimate in another, leading to potential false positives. For example, certain applications rely on frequent timer calls for optimization or benchmarking, which might inadvertently trigger detection mechanisms. Similarly, instrumenting WebAssembly exports introduces performance overhead, which could affect the user experience for non-malicious workloads. Furthermore, the variability of runtime environments, such as differences in browser implementations or hardware configurations, complicates the task of establishing a consistent baseline for timing analysis. Despite these challenges, the combination of timer interception and execution profiling can provide a practical approach for identifying and mitigating time fingerprinting attacks in WebAssembly applications, offering a compromise between enhanced security and acceptable performance.

\subsection{WebAssembly Security}
In recent years, the security implications of WebAssembly have garnered significant attention. Musch et al. \cite{Musch2019} conducted an extensive study on the applications of WebAssembly in real-world environments, categorizing its use into multiple domains such as custom utilities, games, libraries, tests, cryptomining, and obfuscation. Their findings suggest that WebAssembly is likely to serve as a vector for sophisticated web-based malware obfuscation. Subsequently, Hilbig et al. \cite{Hilbig2020} analyzed 8461 unique WebAssembly binaries, revealing that many of them were compiled from memory-unsafe languages such as C and C++. This introduces the potential for traditional memory-related vulnerabilities to propagate into WebAssembly binaries \cite{Lehmann2020}. These studies underscore the dual nature of WebAssembly as a powerful tool for creating high-performance web applications, while simultaneously presenting new challenges for security researchers. The continued evolution of WebAssembly necessitates ongoing research into its security implications to preempt potential misuse in malicious activities.

Lehmann et al. \cite{Lehmann2020} provided a comprehensive analysis of WebAssembly’s linear memory and its utilization by languages such as C, C++, and Rust. They demonstrated that WebAssembly lacks the memory protection features present in native binaries, making it susceptible to various attack vectors, including control flow manipulation and host environment exploitation. To address these concerns, Lehmann et al. \cite{Lehmann2021} proposed \textit{Fuzzm}, the first fuzzer specifically designed for WebAssembly binaries. Using the well-established \textit{AFL fuzzer}, \textit{Fuzzm} identified numerous crashes in real-world WebAssembly binaries and introduced stack and heap canaries to detect memory corruption. These mechanisms act as binary hardening techniques, making WebAssembly binaries more resilient to exploitation.

Cabrera et al. \cite{Cabrera2020} introduced \textit{CROW}, the first code diversification framework for WebAssembly. The authors evaluated \textit{CROW} using a dataset of 303 C programs and successfully diversified 239 of them, demonstrating the tool’s robustness. Moreover, they applied \textit{CROW} to off-the-shelf cryptography libraries such as \textit{libsodium}, showcasing its capability to enhance software diversity and security. In another study, Park et al. \cite{Park2020} developed a cryptography library using WebAssembly, achieving more than a twofold increase in performance over JavaScript. The authors also implemented atomic block-based scalar multiplication to mitigate the risk of side-channel attacks, thereby enhancing the security posture of WebAssembly-based cryptography.

\section{Conclusion}
This paper introduces a novel WebAssembly-based browser fingerprinting method that leverages WebAssembly’s computational capabilities to achieve high accuracy in identifying and differentiating between browsers. By exploiting subtle differences in the WebAssembly JavaScript API implementation, our approach can uniquely identify browsers and devices, even when traditional browser identifiers are completely spoofed. Through extensive evaluation on various physical devices and 158 browser instances—including Google Chrome, Microsoft Edge, Mozilla Firefox, and Safari—our technique demonstrates a false-positive rate of less than 1\% when distinguishing between Chromium-based and non-Chromium-based browsers. Given WebAssembly’s widespread support across all major browsers, this method holds promise for precise user identification, improved security, and more effective personalized experiences. However, recognizing the potential privacy risks, we propose mitigation strategies that introduce minor timing delays to reduce execution variations, thereby diminishing the accuracy of WebAssembly-based fingerprinting. Our findings underscore the need to consider WebAssembly in the context of web privacy and security, providing a foundation for future research and the development of privacy-preserving standards in web-browsers technologies.

\balance
\bibliographystyle{IEEEtran}
\bibliography{IEEEabrv,wa2} 

\begin{thebibliography}{10}
\providecommand{\url}[1]{#1}
\csname url@samestyle\endcsname
\providecommand{\newblock}{\relax}
\providecommand{\bibinfo}[2]{#2}
\providecommand{\BIBentrySTDinterwordspacing}{\spaceskip=0pt\relax}
\providecommand{\BIBentryALTinterwordstretchfactor}{4}
\providecommand{\BIBentryALTinterwordspacing}{\spaceskip=\fontdimen2\font plus
\BIBentryALTinterwordstretchfactor\fontdimen3\font minus
  \fontdimen4\font\relax}
\providecommand{\BIBforeignlanguage}[2]{{%
\expandafter\ifx\csname l@#1\endcsname\relax
\typeout{** WARNING: IEEEtran.bst: No hyphenation pattern has been}%
\typeout{** loaded for the language `#1'. Using the pattern for}%
\typeout{** the default language instead.}%
\else
\language=\csname l@#1\endcsname
\fi
#2}}
\providecommand{\BIBdecl}{\relax}
\BIBdecl

\bibitem{Sirinam2018}
P.~Sirinam, M.~Imani, M.~Juarez, and M.~Wright, ``{Deep fingerprinting:
  Undermining website fingerprinting defenses with deep learning},'' 2018.

\bibitem{Wang2014}
T.~Wang, X.~Cai, R.~Nithyanand, R.~Johnson, and I.~Goldberg, ``{Effective
  attacks and provable defenses for website fingerprinting},'' in
  \emph{Proceedings of the 23rd USENIX Security Symposium}, 2014.

\bibitem{Korczynski2014}
M.~Korczy{\'{n}}ski and A.~Duda, ``{Markov chain fingerprinting to classify
  encrypted traffic},'' in \emph{Proceedings - IEEE INFOCOM}, 2014.

\bibitem{Cao2017}
\BIBentryALTinterwordspacing
Y.~Cao, S.~Li, and E.~Wijmans, ``{(Cross-)Browser Fingerprinting via OS and
  Hardware Level Features},'' in \emph{Proceedings 2017 Network and Distributed
  System Security Symposium}.\hskip 1em plus 0.5em minus 0.4em\relax Reston,
  VA: Internet Society, 2017. [Online]. Available:
  \url{https://www.ndss-symposium.org/ndss2017/ndss-2017-programme/cross-browser-fingerprinting-os-and-hardware-level-features/}
\BIBentrySTDinterwordspacing

\bibitem{Nikiforakis2013}
N.~Nikiforakis, A.~Kapravelos, W.~Joosen, C.~Kruegel, F.~Piessens, and
  G.~Vigna, ``{Cookieless monster: Exploring the ecosystem of web-based device
  fingerprinting},'' in \emph{Proceedings - IEEE Symposium on Security and
  Privacy}, 2013, pp. 541--555.

\bibitem{Fifield2015a}
D.~Fifield and S.~Egelman, ``{Fingerprinting web users through font metrics},''
  in \emph{Lecture Notes in Computer Science (including subseries Lecture Notes
  in Artificial Intelligence and Lecture Notes in Bioinformatics)}, 2015.

\bibitem{Das2017}
A.~Das, N.~Borisov, and M.~Caesar, ``{Tracking Mobile Web Users Through Motion
  Sensors: Attacks and Defenses},'' 2017.

\bibitem{Sanchez-Rola2018}
\BIBentryALTinterwordspacing
I.~Sanchez-Rola, I.~Santos, and D.~B. Eurecom, ``{Clock Around the Clock:
  Time-Based Device Fingerprinting},'' 2018. [Online]. Available:
  \url{https://doi.org/10.1145/3243734.3243796}
\BIBentrySTDinterwordspacing

\bibitem{nmap}
\BIBentryALTinterwordspacing
Fyodor. nmap - free security scanner for network exploration and security
  audits. [Online]. Available: \url{http://nmap.org}
\BIBentrySTDinterwordspacing

\bibitem{Shamsi2014}
Z.~Shamsi, A.~Nandwani, D.~Leonard, and D.~Loguinov, ``{Hershel: Single-packet
  OS fingerprinting},'' in \emph{Performance Evaluation Review}, 2014.

\bibitem{Miettinen2017}
M.~Miettinen, S.~Marchal, I.~Hafeez, N.~Asokan, A.~R. Sadeghi, and S.~Tarkoma,
  ``{IoT SENTINEL: Automated Device-Type Identification for Security
  Enforcement in IoT},'' in \emph{Proceedings - International Conference on
  Distributed Computing Systems}, 2017.

\bibitem{Kumar2018}
\BIBentryALTinterwordspacing
U.~Kumar and S.~Gambhir, ``{Device Fingerprint and Mobile Agent based
  Authentication Technique in Wireless Networks Authentication View project
  Device Fingerprint and Mobile Agent based Authentication Technique in
  Wireless Networks},'' \emph{International Journal of Future Generation
  Communication and Networking}, vol.~11, no.~3, pp. 33--48, 2018. [Online].
  Available: \url{http://dx.doi.org/10.14257/ijfgcn.2018.11.3.04}
\BIBentrySTDinterwordspacing

\bibitem{Zuo2019}
C.~Zuo, Z.~Lin, H.~Wen, and Y.~Zhang, ``{Automatic fingerprinting of vulnerable
  BLE IoT devices with static uuids from mobile apps},'' in \emph{Proceedings
  of the ACM Conference on Computer and Communications Security}, 2019.

\bibitem{Selakovic2016}
\BIBentryALTinterwordspacing
M.~Selakovic and M.~Pradel, ``Performance issues and optimizations in
  javascript: An empirical study,'' in \emph{Proceedings of the 38th
  International Conference on Software Engineering}, ser. ICSE '16.\hskip 1em
  plus 0.5em minus 0.4em\relax New York, NY, USA: Association for Computing
  Machinery, 2016, p. 61–72. [Online]. Available:
  \url{https://doi.org/10.1145/2884781.2884829}
\BIBentrySTDinterwordspacing

\bibitem{Haas2017}
A.~Haas, A.~Rossberg, D.~L. Schuff, B.~L. Titzer, M.~Holman, D.~Gohman,
  L.~Wagner, A.~Zakai, and J.~F. Bastien, ``{Bringing the web up to speed with
  WebAssembly},'' \emph{ACM SIGPLAN Notices}, 2017.

\bibitem{mallik2019man}
A.~Mallik, ``Man-in-the-middle-attack: Understanding in simple words,''
  \emph{Cyberspace: Jurnal Pendidikan Teknologi Informasi}, vol.~2, no.~2, pp.
  109--134, 2019.

\bibitem{mateti2006hacking}
P.~Mateti, ``Hacking techniques in wireless,'' \emph{Handbook of Information
  Security, Threats, Vulnerabilities, Prevention, Detection, and Management},
  vol.~3, p.~83, 2006.

\bibitem{malan2000transport}
G.~R. Malan, D.~Watson, F.~Jahanian, and P.~Howell, ``Transport and application
  protocol scrubbing,'' in \emph{Proceedings IEEE INFOCOM 2000. Conference on
  Computer Communications. Nineteenth Annual Joint Conference of the IEEE
  Computer and Communications Societies (Cat. No. 00CH37064)}, vol.~3.\hskip
  1em plus 0.5em minus 0.4em\relax IEEE, 2000, pp. 1381--1390.

\bibitem{Papapanagiotou2012}
\BIBentryALTinterwordspacing
I.~Papapanagiotou, E.~M. Nahum, and V.~Pappas, ``{Smartphones vs. laptops:
  Comparing web browsing behavior and the implications for caching},'' in
  \emph{Performance Evaluation Review}, vol.~40, no. 1 SPEC. ISS.\hskip 1em
  plus 0.5em minus 0.4em\relax New York, New York, USA: ACM Press, 2012, pp.
  423--424. [Online]. Available:
  \url{http://dl.acm.org/citation.cfm?doid=2254756.2254824}
\BIBentrySTDinterwordspacing

\bibitem{arkin2002remote}
O.~Arkin, ``A remote active os fingerprinting tool using icmp,'' \emph{login:
  the Magazine of USENIX and Sage}, vol.~27, no.~2, pp. 14--19, 2002.

\bibitem{Beck2007}
F.~Beck, O.~Festor, and I.~Chrisment, ``{IPv6 Neighbor Discovery Protocol based
  OS fingerprinting},'' 2007.

\bibitem{Fifield2015b}
\BIBentryALTinterwordspacing
D.~Fifield, A.~Geana, L.~MartinGarcia, M.~Morbitzer, and J.~D. Tygar, ``{Remote
  operating system classification over IPv6},'' in \emph{AISec 2015 -
  Proceedings of the 8th ACM Workshop on Artificial Intelligence and Security,
  co-located with CCS 2015}.\hskip 1em plus 0.5em minus 0.4em\relax New York,
  NY, USA: Association for Computing Machinery, Inc, 10 2015, pp. 57--68.
  [Online]. Available: \url{https://dl.acm.org/doi/10.1145/2808769.2808777}
\BIBentrySTDinterwordspacing

\bibitem{Quynh2010}
N.~A. Quynh, ``{Operating System Fingerprinting for Virtual Machines},''
  \emph{DEF CON 18 Hacking Conference}, 2010.

\bibitem{Kurtz2015}
A.~Kurtz, H.~Gascon, T.~Becker, K.~Rieck, and F.~Freiling, ``{Fingerprinting
  Mobile Devices Using Personalized Configurations},'' \emph{Proceedings on
  Privacy Enhancing Technologies}, 2015.

\bibitem{Zhang2019}
J.~Zhang, A.~R. Beresford, and I.~Sheret, ``{SensorID: Sensor calibration
  fingerprinting for smartphones},'' in \emph{Proceedings - IEEE Symposium on
  Security and Privacy}, vol. 2019-May.\hskip 1em plus 0.5em minus 0.4em\relax
  Institute of Electrical and Electronics Engineers Inc., 5 2019, pp. 638--655.

\bibitem{Olejnik2016}
{\L}.~Olejnik, G.~Acar, C.~Castelluccia, and C.~Diaz, ``{The leaking battery: A
  privacy analysis of the HTML5 battery status API},'' in \emph{Lecture Notes
  in Computer Science (including subseries Lecture Notes in Artificial
  Intelligence and Lecture Notes in Bioinformatics)}, 2016.

\bibitem{Aksoy2017}
A.~Aksoy, S.~Louis, and M.~H. Gunes, ``{Operating system fingerprinting via
  automated network traffic analysis},'' in \emph{2017 IEEE Congress on
  Evolutionary Computation, CEC 2017 - Proceedings}, 2017.

\bibitem{Matsunaka2013}
T.~Matsunaka, A.~Yamada, and A.~Kubota, ``{Passive OS fingerprinting by DNS
  traffic analysis},'' in \emph{Proceedings - International Conference on
  Advanced Information Networking and Applications, AINA}, 2013.

\bibitem{Lastovicka2018}
M.~Lastovicka, T.~Jirsik, P.~Celeda, S.~Spacek, and D.~Filakovsky, ``{Passive
  os fingerprinting methods in the jungle of wireless networks},'' in
  \emph{IEEE/IFIP Network Operations and Management Symposium: Cognitive
  Management in a Cyber World, NOMS 2018}, 2018.

\bibitem{Lastovicka2020}
M.~Lastovicka, S.~Spacek, P.~Velan, and P.~Celeda, ``{Using TLS Fingerprints
  for OS Identification in Encrypted Traffic},'' in \emph{Proceedings of
  IEEE/IFIP Network Operations and Management Symposium 2020: Management in the
  Age of Softwarization and Artificial Intelligence, NOMS 2020}.\hskip 1em plus
  0.5em minus 0.4em\relax Institute of Electrical and Electronics Engineers
  Inc., 4 2020.

\bibitem{Chen2014}
Y.~C. Chen, Y.~Liao, M.~Baldi, S.~J. Lee, and L.~Qiu, ``{OS fingerprinting and
  tethering detection in mobile networks},'' in \emph{Proceedings of the ACM
  SIGCOMM Internet Measurement Conference, IMC}, 2014.

\bibitem{Mowery2012}
K.~Mowery and H.~Shacham, ``{Pixel Perfect : Fingerprinting Canvas in HTML5},''
  \emph{Web 2.0 Security {\&} Privacy 20 (W2SP)}, 2012.

\bibitem{Mowery2011}
K.~Mowery, D.~Bogenreif, S.~Yilek, and H.~Shacham, ``{Fingerprinting
  Information in JavaScript Implementations},'' \emph{Web 2.0 Security {\&}
  Privacy}, 2011.

\bibitem{Schwarz2019}
M.~Schwarz, F.~Lackner, and D.~Gruss, ``{JavaScript Template Attacks:
  Automatically Inferring Host Information for Targeted Exploits},'' no.
  February, 2019.

\bibitem{Queiroz2019}
\BIBentryALTinterwordspacing
J.~S. Queiroz and E.~L. Feitosa, ``{A Web Browser Fingerprinting Method Based
  on the Web Audio API},'' \emph{The Computer Journal}, vol.~62, no.~8, pp.
  1106--1120, 8 2019. [Online]. Available:
  \url{https://academic.oup.com/comjnl/article/62/8/1106/5298776}
\BIBentrySTDinterwordspacing

\bibitem{Franklin2008}
J.~Franklin, M.~Luk, J.~M. McCune, A.~Seshadri, A.~Perrig, and L.~{Van Doorn},
  ``{Remote detection of virtual machine monitors with fuzzy benchmarking},''
  in \emph{Operating Systems Review (ACM)}, vol.~42, no.~3, 4 2008, pp. 83--92.

\bibitem{Ho2014}
G.~Ho, D.~Boneh, L.~Ballard, and N.~Provos, ``{Tick tock: Building browser red
  pills from timing side channels},'' \emph{8th USENIX Workshop on Offensive
  Technologies, WOOT 2014}, 2014.

\bibitem{Gu2012}
\BIBentryALTinterwordspacing
Y.~Gu, Y.~Fu, A.~Prakash, Z.~Lin, and H.~Yin, ``{OS-SOMMELIER: Memory-only
  operating system fingerprinting in the cloud},'' in \emph{Proceedings of the
  3rd ACM Symposium on Cloud Computing, SoCC 2012}.\hskip 1em plus 0.5em minus
  0.4em\relax New York, New York, USA: ACM Press, 2012, pp. 1--13. [Online].
  Available: \url{http://dl.acm.org/citation.cfm?doid=2391229.2391234}
\BIBentrySTDinterwordspacing

\bibitem{Owens2011}
R.~Owens and W.~Wang, ``{Non-interactive OS fingerprinting through memory
  de-duplication technique in virtual machines},'' in \emph{Conference
  Proceedings of the IEEE International Performance, Computing, and
  Communications Conference}, 2011.

\bibitem{Husak2016}
M.~Hus{\'{a}}k, M.~{\v{C}}erm{\'{a}}k, T.~Jirs{\'{i}}k, and P.~{\v{C}}eleda,
  ``{HTTPS traffic analysis and client identification using passive SSL/TLS
  fingerprinting},'' \emph{Eurasip Journal on Information Security}, 2016.

\bibitem{Zhou2014}
Z.~Zhou, W.~Diao, X.~Liu, and K.~Zhang, ``{Acoustic fingerprinting revisited:
  Generate stable device ID stealthily with inaudible sound},'' in
  \emph{Proceedings of the ACM Conference on Computer and Communications
  Security}, 2014.

\bibitem{Das2014}
A.~Das, N.~Borisov, and M.~Caesar, ``{Do you hear what i hear? Fingerprinting
  smart devices through embedded acoustic components},'' in \emph{Proceedings
  of the ACM Conference on Computer and Communications Security}, 2014.

\bibitem{Chen2020}
\BIBentryALTinterwordspacing
J.~Chen, K.~He, J.~Chen, Y.~Fang, and R.~Du, ``{PowerPrint: Identifying
  Smartphones through Power Consumption of the Battery},'' 2020. [Online].
  Available: \url{https://doi.org/10.1155/2020/3893106}
\BIBentrySTDinterwordspacing

\bibitem{Takei2015}
N.~Takei, T.~Saito, K.~Takasu, and T.~Yamada, ``{Web Browser Fingerprinting
  Using only Cascading Style Sheets},'' in \emph{Proceedings - 2015 10th
  International Conference on Broadband and Wireless Computing, Communication
  and Applications, BWCCA 2015}, 2015.

\bibitem{Shusterman2018}
A.~Shusterman, L.~Kang, Y.~Haskal, Y.~Meltser, P.~Mittal, Y.~Oren, and
  Y.~Yarom, ``{Robust website fingerprinting through the cache occupancy
  channel},'' 2018.

\bibitem{Emscripten}
\BIBentryALTinterwordspacing
Emscripten. Emscripten is a complete compiler toolchain to webassembly.
  [Online]. Available: \url{https://emscripten.org}
\BIBentrySTDinterwordspacing

\bibitem{LLVM:CGO04}
C.~Lattner and V.~Adve, ``{LLVM}: A compilation framework for lifelong program
  analysis and transformation,'' in \emph{CGO}, San Jose, CA, USA, 3 2004, pp.
  75--88.

\bibitem{SDL}
\BIBentryALTinterwordspacing
SDL. Sdl. [Online]. Available: \url{https://www.libsdl.org/}
\BIBentrySTDinterwordspacing

\bibitem{pthreads}
\BIBentryALTinterwordspacing
man7. pthreads(7). [Online]. Available:
  \url{https://man7.org/linux/man-pages/man7/pthreads.7.html}
\BIBentrySTDinterwordspacing

\bibitem{asmjs}
\BIBentryALTinterwordspacing
asmjs. asm.js - frequently asked questions. [Online]. Available:
  \url{http://asmjs.org/faq.html}
\BIBentrySTDinterwordspacing

\bibitem{Zakai2017}
\BIBentryALTinterwordspacing
A.~Zakai. Why webassembly is faster than asm.js. [Online]. Available:
  \url{https://hacks.mozilla.org/2017/03/why-webassembly-is-faster-than-asm-js/}
\BIBentrySTDinterwordspacing

\bibitem{WebAssemblyMajorBrowsers}
\BIBentryALTinterwordspacing
J.~McConnell. Webassembly support now shipping in all major browsers. [Online].
  Available:
  \url{https://blog.mozilla.org/blog/2017/11/13/webassembly-in-browsers}
\BIBentrySTDinterwordspacing

\bibitem{Musch2019}
M.~Musch, C.~Wressnegger, M.~Johns, and K.~Rieck, ``{New kid on the web: A
  study on the prevalence of webassembly in the wild},'' in \emph{Lecture Notes
  in Computer Science (including subseries Lecture Notes in Artificial
  Intelligence and Lecture Notes in Bioinformatics)}, 2019.

\bibitem{Hilbig2020}
\BIBentryALTinterwordspacing
A.~Hilbig, D.~Lehmann, and M.~Pradel, ``An empirical study of real-world
  webassembly binaries security, languages, use cases ccs concepts • security
  and privacy → software and application security. acm reference format,''
  p.~13, 2020, authors have collected a dataset of 8461 unique WebAssembly
  binaries. They concluded that the majority of binaries from memory-unsafe
  languages,. [Online]. Available:
  \url{https://doi.org/10.1145/3442381.3450138}
\BIBentrySTDinterwordspacing

\bibitem{Lehmann2020}
\BIBentryALTinterwordspacing
D.~Lehmann, J.~Kinder, and M.~Pradel, ``Everything old is new again: Binary
  security of webassembly,'' 2020, they performed in-depth analysis of Wasm's
  linear memory and its use by programs compiled from languages such a C/C++
  and Rust. They show that memory protections are missing from Wasm, making
  wasm binaries less secured than their binaries counterpart. [Online].
  Available:
  \url{https://www.usenix.org/conference/usenixsecurity20/presentation/lehmann}
\BIBentrySTDinterwordspacing

\bibitem{Lehmann2021}
\BIBentryALTinterwordspacing
D.~Lehmann, M.~T. Torp, and M.~Pradel, ``Fuzzm: Finding memory bugs through
  binary-only instrumentation and fuzzing of webassembly,'' 2021, they
  introduce Fuzzm, the first wasm binarry fuzzer, which uses the popular AFL
  fuzzer. In addition, they also itroduce stack and heap canaries to detect
  overflows and underflows. They showed that fuzzm finds a substantial amount
  of crashes in real-world wasm binarries, while being efficient enough to
  perform hunderds od executions per second. The canaries also serve as a
  stand-alone binary hardening technique to prevent exploitation of vulnerable
  binaries in production. [Online]. Available: \url{https://wasmtime.dev/}
\BIBentrySTDinterwordspacing

\bibitem{Cabrera2020}
\BIBentryALTinterwordspacing
J.~C. Arteaga, O.~Floros, O.~V. Perez, B.~Baudry, and M.~Monperrus, ``Crow:
  Code diversification for webassembly,'' 2020, in this paper they presented
  CROW, the first code diversification approach for WebAssembly. They evaluated
  CROW's capabilities on 303 C programs. CROW was able to diversify 239 of
  them. It was also able to diversify off-the-shelf cryptographic software
  (libsoduim). [Online]. Available:
  \url{https://dx.doi.org/10.14722/madweb.2021.23xxx}
\BIBentrySTDinterwordspacing

\bibitem{Park2020}
\BIBentryALTinterwordspacing
B.~Park, J.~Song, and S.~C. Seo, ``Efficient implementation of a crypto library
  using web assembly,'' \emph{Electronics 2020, Vol. 9, Page 1839}, vol.~9, p.
  1839, 11 2020, they implemented a cryptographic library in WebAssembly. The
  proposed library showed more than 2 times performance improvement in
  WebAssembly compared to JavaScript. In addition, they intoduce atomic
  block-based scalar multiplication, which provides enhanced perforamnce and
  resistance against SCA. [Online]. Available:
  \url{https://www.mdpi.com/2079-9292/9/11/1839/htm
  https://www.mdpi.com/2079-9292/9/11/1839}
\BIBentrySTDinterwordspacing

\end{thebibliography}

\end{document}